\documentclass[two column]{aastex631}

\received{March 7, 2024}
\revised{April 9, 2024}
\accepted{April 12, 2024}

\shorttitle{Searching for Hyper-compact star clusters in the Milky Way using LAMOST and {\it Gaia}}
\shortauthors{Wu et al.}

\graphicspath{{./}{figures/}}

\begin{document}

\title{Searching for hyper-compact star clusters in the Milky Way using LAMOST and {\it Gaia}} 
\author[0009-0006-7193-4443]{Hao Wu}
\affiliation{
Institute for Frontiers in Astronomy and Astrophysics, Beijing Normal University,  Beijing 102206, China}
\affiliation{Department of Astronomy, Beijing Normal University, Beijing 100871,China}
\affiliation{Department of Astronomy, Peking University, Beijing 100871, China}
\affiliation{Kavli Institute for Astronomy and Astrophysics, Peking University, Beijing 100871, China}

\author[0000-0003-2471-2363]{Haibo Yuan}
\affiliation{
Institute for Frontiers in Astronomy and Astrophysics, Beijing Normal University,  Beijing 102206, China}
\affiliation{Department of Astronomy, Beijing Normal University, Beijing 100871,China}

\author[0000-0002-3530-3277]{Yilun Wang}
\affiliation{National Astronomical Observatories, Chinese Academy of Sciences, Beijing 100101, China}
\affiliation{School of Astronomy and Space Sciences, University of Chinese Academy of Sciences, Beijing 100049, China}

\author[0000-0002-3651-0681]{Zexi Niu}
\affiliation{National Astronomical Observatories, Chinese Academy of Sciences, Beijing 100101, China}
\affiliation{School of Astronomy and Space Sciences, University of Chinese Academy of Sciences, Beijing 100049, China}

\author[0000-0002-7727-1699]{Huawei Zhang}
\affiliation{Department of Astronomy, Peking University, Beijing 100871, China}
\affiliation{Kavli Institute for Astronomy and Astrophysics, Peking University, Beijing 100871, China}

\correspondingauthor{Haibo Yuan email:yuanhb@bnu.edu.cn}
\begin{abstract}
During the early merger of the Milky Way, intermediate-mass black holes in merged dwarf galaxies may have been ejected from the center of their host galaxies due to gravitational waves, carrying some central stars along. This process can lead to the formation of hyper-compact star clusters, potentially hosting black holes in the mass range of $10^4$ to $10^5$ solar masses. These clusters are crucial targets for identifying and investigating intermediate-mass black holes. However, no hyper-compact star clusters in the Milky Way have been identified so far.
In this paper, taking advantage of the high spatial resolution power of {\it Gaia}, we used data from {\it Gaia} EDR3 and LAMOST DR7, along with additional data from Pan-STARRS and SDSS, to conduct an initial screening of 6,138,049 sources using various parameters of Gaia EDR3. A total of 4,786 sources were selected for in-depth analysis. Each of these sources was meticulously scrutinized by examining their images, spectra, and nearby celestial objects to exclude various false positives, such as contaminations, galaxies, wide binaries, or wrong matches. 
We finally identified one likely hyper-compact star cluster candidate in the Milky Way, laying the foundation for further high-resolution imaging and spectral verification.
\end{abstract}

\section{Introduction}\label{introduction}

Gravitational waves (GWs) are predicted to be emitted during the coalescence of black holes (BHs) and carry away momentum flux, causing the remnant BHs to recoil in the opposite direction (e.g. \citealt{Peres1962,Bekenstein1973,Fitchett1983,Redmount1989}). This process resembles the remnant BHs being ``kicked", with velocities numerically calculated ranging from $\sim$ 100 to thousands of $\rm{km\,s}^{-1}$, depending on factors such as spinning, mass ratio, and orientation of the binary BHs, and typically less than 500 $\rm{km\,s}^{-1}$ (e.g., \citealt{Favata2004,Merritt2004,Campanelli2007,Lousto2011,Lousto2012}). With such typical kicking velocities, intermediate-mass BHs (ranging from $10^3$ to $10^5$ solar masses) residing in low-mass galaxies with shallower potential-wells and lower escape velocities are more likely to be ejected from their host galaxies \citep{o2009star}. However, these kicking velocities are smaller than the escape velocity of the Milky Way (MW). Therefore, numerous remnant BHs are predicted to be kicked out of the low-mass galaxies during the formation of the MW and should be floating in the MW halo today \citep{o2009star}. Each black hole carries a number of stars that used to reside in the central region of the host galaxy, bound to the black hole \citep{Komossa2008}. The size for such systems is predicted to be related to numerous factors, such as the central stellar velocity dispersion of the host galaxy and the nucleus dynamics (\citealt{Merritt2009,lena2020hypercompact}), and more likely to be smaller than 1 parsec (\citealt{Merritt2009,o2009star,greene2021search}), leading to Hyper-compact star clusters (HCSCs, \citealt{Merritt2009,o2009star,lena2020hypercompact}). It is estimated that there may be approximately hundreds of HCSCs in the MW halo today, each potentially carrying an intermediate-mass black hole (\citealt{libeskind2006effect,o2009star}). 

Identifying HCSCs would not only confirm the occurrence of black hole mergers but also showcase the effectiveness of GWs in expelling the remnant BHs from their host galaxies \citep{lena2020hypercompact}. This may explain the absence of BHs in certain dwarf galaxies and globular clusters \citep{Merritt2004}. Additionally, a comprehensive characterization of these HCSCs, including measurements of velocity dispersion, stellar mass, and effective radius, holds the potential to provide valuable insights into the distribution of GW recoils and foster a deeper comprehension of host galaxy formation dynamics, as well as the distribution of stars in the core region during mergers \citep{lena2020hypercompact}. 

Due to the significant potential of identifying HCSCs and the intermediate-mass BHs possibly associated with, several studies have been conducted, providing guidance for identification. \cite{o2009star} calculated the dynamical evolution of HCSCs, predicted their size and argued that existing ground-based photometric surveys are unable to resolve individual stars located in the centers. \cite{o2012recoiled} conducted long-term N-body simulations to determine the stellar distribution of HCSCs. Their findings suggested that 20\% to 90\% of stars of HCSCs may have been removed within about 1 billion years, due to scattering and tidal disruption. They also analyze the photometric and spectroscopic properties of the remaining stars in HCSCs. In light of these results, they attempted to search for HCSCs in the MW using Sloan Digital Sky Survey Data Release 7 \citep[SDSS DR7;][] {Abazajian2009}. As a result, although $\sim$ 100 HCSC candidates had been identified by \cite{o2012recoiled}, none of them has been validated. \cite{lena2020hypercompact} provided reliable photometric and spectral characteristics of HCSCs through simulated observations. They simulated images of HCSCs with different ages, metallicities, distances, and kicking velocities in the Pan-STARRS \citep{chambers2016pan} and Euclid \citep{laureijs2011euclid} data. Additionally, \cite{lena2020hypercompact} simulated the broad-band spectra and corresponding multi-band colors of HCSCs. They found that the majority of simulated HCSCs' locations in the color-color diagram were consistent with those of stars and galaxies, with spectra resembling K giants but exhibit blue excess.
More recently, \cite{greene2021search} predicted that HCSCs have a size of approximately 0.5--1 parsec and contain 500--5000 stars and are expected to exist within a range of 30--50 kilo-parsec from the Galactic center. Based on these predictions, Early Data Release 3 of the {\it Gaia} mission ({\it Gaia} EDR3; \citealt{gaia2016gaia,Gaiadr2,Gaiaedr3}) and Dark Energy Camera Legacy Survey \citep[DECaLS;][] {dey2019overview} data within an area exceeding 8000 square degrees were utilized to derive the stellar counts around each star in {\it Gaia} EDR3, using a Negative Binomial model. They further investigated significant outliers with large stellar counts to identify HCSCs in the MW. To date, however, no HCSCs in the MW have been identified.

{\it Gaia} satellite scans the entire sky with a high spatial resolution comparable to that of the Hubble Space Telescope \citep{gaia2016gaia}. Moreover, high precision astrometric measurements, along with numerous astrometric parameters provided by {\it Gaia}, make it highly promising to search for extended sources like clusters using the data released by {\it Gaia}.
For instance, \cite{voggel2020gaia} employed stringent parameter thresholds with {\it Gaia} Data Release 2 ({\it Gaia} DR2; \citealt{Gaiadr2}) to identify 632 new luminous clusters candidates situated within the halo of the galaxy Centaurus A. Similarly, \cite{Wang2023} effectively harnessed data sourced from {\it Gaia} EDR3 in conjunction with Pan-STARRS1 DR1, resulting in the successful discovery of 50 novel candidates for globular clusters (GCs) within the Andromeda galaxy (M\,31). 
Meanwhile, The Large-Area Multi-Object Fiber Optic Spectroscopic Telescope \citep[LAMOST;][] {cui2012large} stands out as a telescope with a high rate of spectral acquisition and has assembled one of the largest spectral datasets of sources in the Galaxy \citep{yan2022overview}. 
LAMOST data is very helpful in excluding contaminations from compact extragalactic sources via  redshift measurements. 
Combining Gaia and LAMOST, it is possible to select compact yet extended sources in the Galaxy, i.e., HCSC candidates.

In this work, we conduct a comprehensive exploration for HSCSs in the MW using the {\it Gaia} EDR3 and the LAMOST DR7 data. The data and method used are described in Sections~\ref{section:data} and ~\ref{section:process}, respectively. The results are presented in Section~\ref{section:result}. We summarize and conclude in Section~\ref{section:summary_discussion}.

\section{Data}\label{section:data}

In this study, we used the LAMOST DR7 LRS dataset. Initially, we conducted a cross-matching process with the {\it Gaia} EDR3 within a 1 arcsec radius, resulting in a total of 6,138,049 sources, each associated with corresponding LAMOST DR7 spectra. Additionally, data from Pan-STARRS DR1 and SDSS DR15 were incorporated to assist in the screening process, as detailed in the Section~\ref{section:process}. Detailed descriptions of each of these catalogs are provided below.

\subsection{{\it Gaia} EDR3}\label{section:data_gaia}

The {\it Gaia} satellite, launched by the European Space Agency (ESA) in 2013, serves as a space observatory with the primary mission of constructing a three-dimensional map of our MW galaxy \citep{turon2005three}. {\it Gaia}'s primary goal is to precisely measure the positions, distances, and motions of over a billion stars in the MW, offering insights into the structure, composition, and evolution of the MW \citep{lindegren2007gaia}.

The unique CCD of {\it Gaia} enables the separation of extended sources from normal single stars. The Astrometry Field (AF) is the primary part of the CCD, which collects astrometry and $G$-band photometry information. {\it Gaia}'s CCDs possess along-scan (AL) and across-scan (AC) directions. The instantaneous spatial resolution along the AL axis is comparable to that of the HST \citep{gaia2016gaia}. When an object traverses the {\it Gaia} satellite's CCDs, it is encompassed by a window determined by its brightness. The window size in AL direction is 0\arcsec.7 (equivalent to 12 pixels) for objects falling within the magnitude range of $16{<}G{<}20.7$ and will increase to 1\arcsec.0 (18 pixels) for objects with $13{<}G{<}16$. In this window, the data undergoes a fitting process using a line spread function (LSF), ultimately resulting in the extraction of both the centroid position and flux measurements ($G$-band photometry) \citep{lindegren2018gaia,rowell2021gaia}. Notably, each objects falling in the {\it Gaia}'s field is treated as a single star \citep{rowell2021gaia}. For extended sources, especially those with angular diameter larger than 0\arcsec.7, the use of a fixed window and LSF fitting method may lead to fainter $G$ magnitudes than they should be. On the other hand, the magnitudes in the other two passbands, $BP$ and $RP$, covering smaller wavelength ranges from approximately 330 to 680 nm, and 630 to 1050 nm, respectively \citep{Jordi2010}, are aperture magnitudes derived from a more broader aperture covering 60 pixels along the AL direction, capturing more flux compared to $G$ band. In addition to magnitudes in the $G/BP/RP$ band, we use the following parameters to aid in the selection of HCSC candidates.

\begin{enumerate}
    \item BRExcess: phot\_bp\_rp\_excess\_factor, indicating the excess flux of $BP$ and $RP$ with respect to $G$ \citep{riello2021gaia}. For extended sources, the measured G-band flux would be smaller than their actual flux, and the BRExcess value will be larger. 
    
    \item AEN: astrometric\_excess\_noise, measuring the difference between the best-fitting astrometric model for single point sources and the observations of a source (\citealt{Gaiadr2,astrometric_core_solution}). For extended sources like clusters, this parameter should be significant larger than normal single star.
    
    \item RUWE: Astrometric renormalised unit weight error of a source. Sources with a RUWE value significantly surpassing 1.0 may be non-single or it is challenging to derive the astrometric solution of them \citep{pourbaix2022gaia}. Additionally, sources with only a two-parameter solution would lack RUWE value. \citep{hobbs2022gaia}.
    
    \item IGHA: ipd\_gof\_harmonic\_amplitude, indicating the variation amplitude of the Image Parameter Determination (IPD) Goodness of Fit (GoF) relative to the position angle of the scan direction \citep{hobbs2022gaia}. The larger the parameter, the more likely the source is to have an asymmetric structure (e.g., binary star/ galaxy).
    
    \item IFMP: ipd\_frac\_mumlti\_peak, which is the percentage of successful IPD windows with more than one peak \citep{hobbs2022gaia}. It characterizes situations where the target source exhibits multiple peaks within the {\it Gaia} photometric radius. This parameter allows us to examine the photometric distribution and morphological information of the source.

    \item phot\_bp(/rp)\_n\_obs: This parameter describes the number of observations that contribute to the mean $BP$(/$RP$) band flux and the corresponding flux error of the target source \citep{riello2021gaia}.
    
    \item phot\_bp(/rp)\_n\_blended\_transits: This parameter characterizes the number of transits evaluated as blended among those contributing to the mean spectrum. A transit is classified as blended when more than one source is within the window \citep{de2022gaia}. Combining this parameter with phot\_bp(/rp)\_n\_contaminated\_transits, we can assess the reliability of the $BP$(/$RP$) spectra \citep{de2022gaia}.
    
    \item phot\_bp(/rp)\_n\_contaminted\_transits: Similar to phot\_bp(/rp)\_n\_blended\_transits parameter, but quantifies the transits evaluated as contaminated. A transit is deemed contaminated when a portion of the flux within the window is estimated to come from a nearby source situated outside the window \citep{de2022gaia}. Combining this parameter with phot\_bp(/rp)\_n\_blended\_transits enables the assessment of the reliability of the $BP$ (/$RP$) spectra \citep{de2022gaia}.
\end{enumerate}
\subsection{LAMOST DR7 LRS}\label{section:data_lamost}
LAMOST, also known as the Guo Shou Jing Telescope, is a big astronomical facility located in Xinglong, China. It combines a wide field of view of 20 square degrees and a large effective aperture of 4 meters \citep{cui2012large}. In its low-resolution observing mode (resolution R $\approx$ 1800), LAMOST can simultaneously control 4000 optical fibers on a 1.75-meter diameter focal plane. These fibers are connected to 32 CCDs, covering a wavelength range of 370--900 nm, with a limiting magnitude of approximately 18.
By the end of 2022, LAMOST has accumulated more than 21 million spectra, consisting of 11 million low resolution spectra and 10 million medium resolution spectra \citep{yan2022overview}.

In this study, we used the data obtained from the LAMOST DR7 LRS. In addition to spectra, this data release incorporates stellar parameters determined by the LAMOST Stellar Parameter Pipeline \citep[LASP;][]{luo2015first}, a specialized program designed for LAMOST to perform fully automated spectrum analysis using stellar template matching techniques. For objects classified as stars, the stellar parameters such as effective temperature, surface gravity, metallicity and radial velocity are provided.  

\subsection{Pan-STARRS DR1}\label{section:data_PS1}
Located in Hawaii, the Panoramic Survey Telescope and Rapid Response System (Pan-STARRS) has conducted a comprehensive $3\pi$ Steradian Survey employing five broad-band filters, denoted as $grizy_{p1}$. Within these bands, the $5\sigma$ magnitude limits for the faint end are $(23.3, 23.2, 23.1, 22.3, 21.4)$ mag, while for the bright end, they are $(14.5, 15.0, 15.0, 14.0, 13.0)$ mag \citep{chambers2016pan}. In our study, we used both the point spread function ($PSF$) magnitudes and the aperture magnitudes. The PS1 images were also used in this work.

\subsection{Sloan Digital Sky Survey DR15}\label{section:data_SDSS}

The Sloan Digital Sky Survey (SDSS) employs a 2.5-meter wide-field telescope equipped with $ugriz$ photometric filters. SDSS Data Release 15 \citep[SDSS DR15;][]{aguado2019fifteenth} encompasses observed objects along with their associated parameters, including classifications (point source or extended source) and magnitudes in $ugriz$ bands \citep{eisenstein2011sdss}. Color images for each object in SDSS DR15 are also available. Additionally, a subset of objects in SDSS DR15 has undergone spectroscopic observations, resulting in corresponding spectra.

\section{Methods}\label{section:process}
Our methods include the classification of the sample based on the BRExcess parameter to select sources with significantly large BRExcess values for in-depth analysis. Subsequently, we employ 6 additional parameters and establish empirical criteria for these parameters to select HCSC candidates and eliminate various false positive cases within these sources.

\subsection{Classification based on BRExcess}\label{section:BRExcess}

\begin{figure}[htbp!] 
\centering
\includegraphics[scale=0.8]{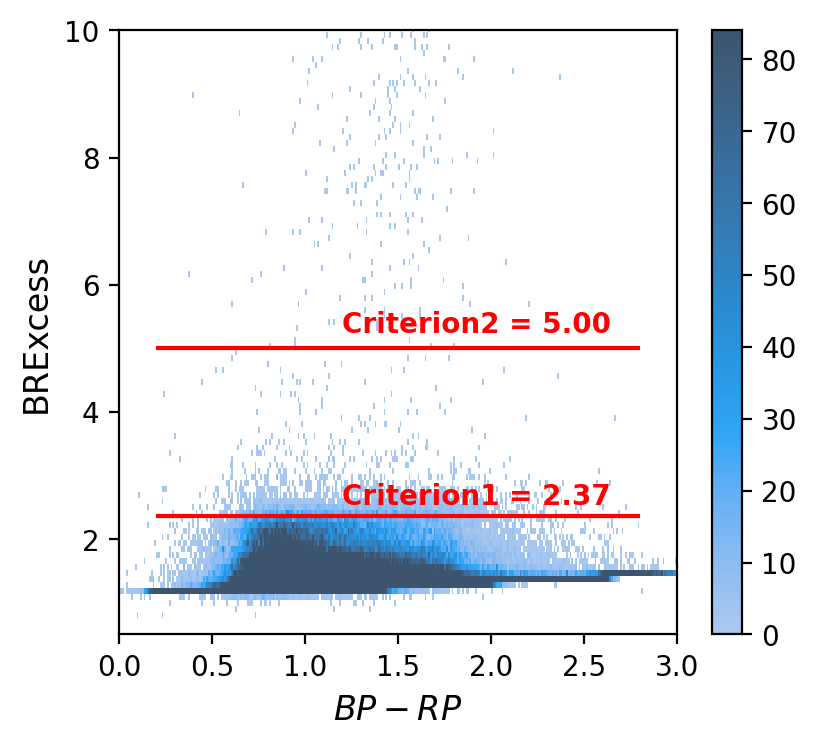}
\caption{Distribution of BRExcess as a function of $BP - RP$ for all 6,138,049 sources. The colorbar indicates the number density. The sample is divided into three sub-samples by the two red lines.}
\label{fig:xfig1}
\end{figure}

In this work, we firstly classify the 6,138,049 sources using the BRExcess parameter. Figure\,\ref{fig:xfig1} shows the distribution of BRExcess in relation to $BP - RP$ for all sources. We establish two criteria to divide the sample into three sub-samples: 701 sources with 5 $<$ BRExcess, 4,085 sources with 2.37 $< BRExcess \leq$ 5, and the others with BRExcess $\leq$ 2.37. Given that most sources with BRExcess $\leq$ 2.37 are more likely point sources, our focus is on the two sub-samples with $\text{BRExcess} > 2.37$.

\subsection{Further classification by other {\it Gaia} EDR3 parameters}\label{section:parameters}

\begin{figure*}[htbp!] 
\centering
\includegraphics[scale=0.75]{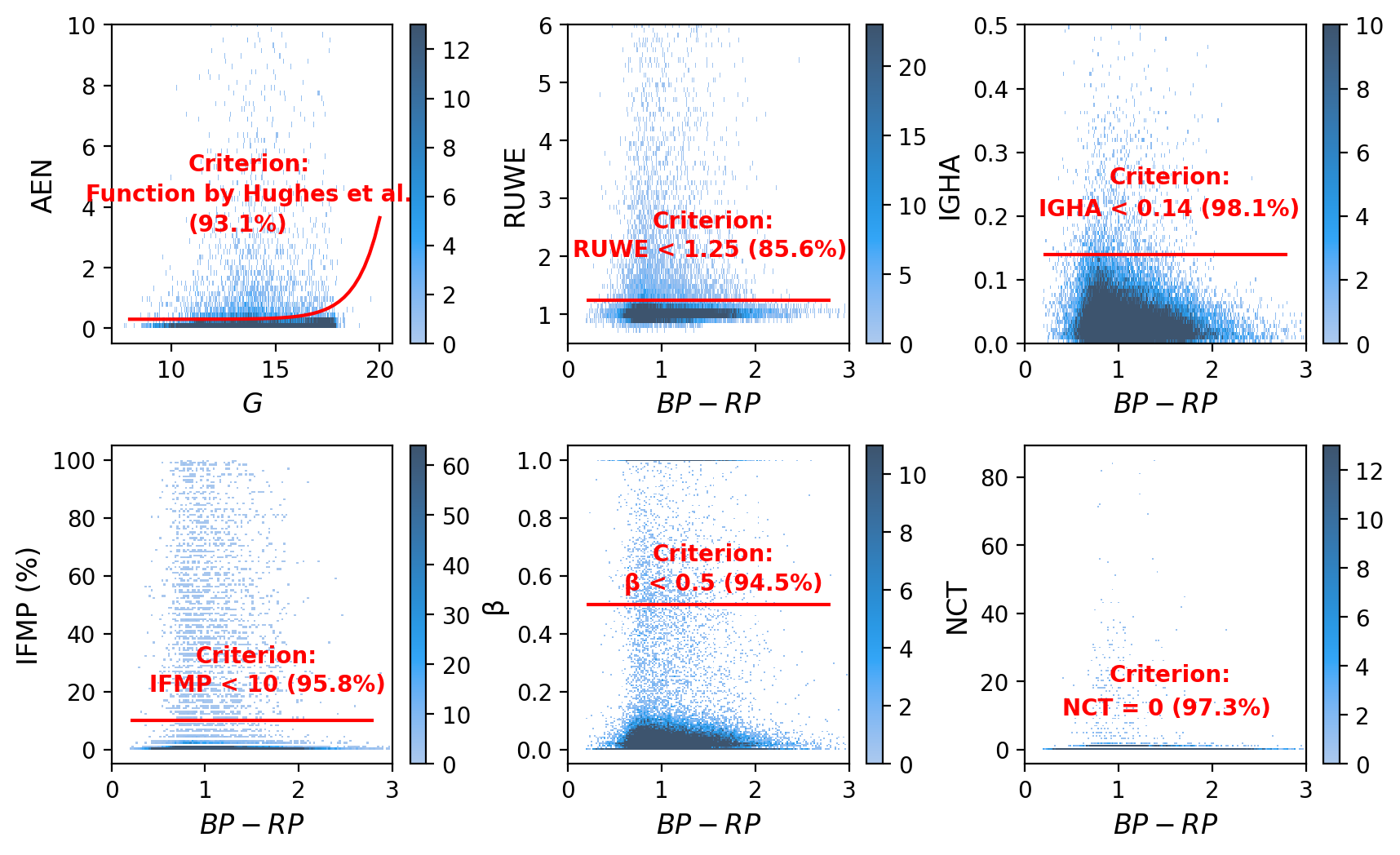}
\caption{Distribution of the six {\it Gaia} EDR3 parameters relative to $G$ or $BP - RP$ for a random sample of 60,671 LAMOST sources with BRExcess $\leq$ 2.37. In each panel, the red line and accompanying text represent the selection criteria employed in this study. Additionally, the percentages displayed indicate the proportion of randomly sampled LAMOST sources that fall below the selection criteria.}
\label{fig:xfig2}
\end{figure*}

To characterize the features of HCSC candidates and further distinguish them from other sources, such as normal single stars or binaries, we randomly sample 60,671 LAMOST sources with BRExcess $\leq$ 2.37 from those 6,138,049 sources. Figure\,\ref{fig:xfig2} plots their {\it Gaia} EDR3 parameters as a function of $G$ or $BP-RP$. In each panel, the red line, along with accompanying annotations, illustrates the selection criteria for HCSC employed in this study. 

\begin{enumerate}
    \item AEN: \cite{voggel2020gaia} used AEN along with BRExcess to identify over 600 new cluster candidates in the halo of NGC\,5128. In a subsequent study, \cite{hughes2021ngc} employed this method to identify a total of 1450 $\pm$ 160 GC candidates in NGC\,5128. Using AEN, they defined a line $3\sigma$ above the average foreground stellar's AEN value, as shown in Equation~\ref{Eq.3-1}. \cite{obasi2023globular} applied this criterion to search for GCs in the Circinus galaxy. They validated the criterion with identified GCs in M\,81 and NGC\,5128, concluding that all GCs within a distance of $\pm$ 10 kilo-parsec from the galaxies' center satisfied the criterion. Ultimately, \cite{obasi2023globular} obtained a sample of 78 potential GCs in the Circinus galaxy. In our study, we also tested 453 confirmed GCs in M\,31 \citep{wang2014new} against this criterion, and found that 96.7\% of them have ${\rm AEN_{source}}$ \textgreater ~${\rm AEN_{3\sigma}}$. 
    \begin{equation}
    \label{Eq.3-1}
    {\rm AEN_{3\sigma}}=0.297+5.36\times10^{-8}e^{0.895G} 
    \end{equation}
    Since the aim of our study is to search for HCSCs, which are also compact cluster systems, we utilize the aforementioned AEN criterion. If a source has ${\rm AEN_{source}}$ \textgreater ~${\rm AEN_{3\sigma}}$, it is more likely to be classified as an extended source, denoted as ``abnormal". Conversely, the source is considered to be more likely a point source (such as a single star), denoted as ``normal". The top left panel of Figure\,\ref{fig:xfig2} shows the distribution of the AEN value with $G$ magnitude of the 60,671 LAMOST sources with BRExcess $\leq$ 2.37, indicating that 93.1\% of these sources exhibit ${\rm AEN_{source}}$ $\leq$ ${\rm AEN_{3\sigma}}$, which aligns with our expectations.

    \item RUWE: 
    Studies have established RUWE $\geq$ 1.25 as a criterion to select binary stars (e.g. \citealt{Andrew2022,Penoyre2022}), which is consistent with our inspection shown in Figure\,\ref{fig:xfig2}. To avoid the contamination of binary systems, we define sources with RUWE $\geq$ 1.25  to be ``abnormal". 
    Conversely, sources with RUWE $<$ 1.25 are classified as ``normal".
    \item IGHA: The observed HCSCs are expected to exhibit structural symmetry \citep{lena2020hypercompact}. We define sources with IGHA $\geq$ 0.14 as ``abnormal", while sources with IGHA $<$ 0.14 as ``normal".
    In Section~\ref{section:2.37_brexcess_5}, we also set relatively lenient or stringent screening criteria for this parameter in order to retain more possible candidates for further validation.
    \item IFMP: \cite{mannucci2022unveiling} effectively utilized the IFMP parameter to identify several binary quasar candidates. We define sources with IFMP $\geq$ 10 as ``abnormal", indicating a high likelihood of being binary systems, while sources with IFMP $<$ 10  as ``normal". 
    In Section~\ref{section:2.37_brexcess_5}, we also apply relatively lenient or stringent screening criteria for this parameter to encompass a wider spectrum of potential candidates.
    \item phot\_bp(/rp)\_n\_blended\_transits: The blending fraction $\beta$ is defined by Equation~\ref{eq:1} \citep{riello2021gaia}, where A represents phot\_bp\_n\_blended\_transits, B represents phot\_rp\_n\_blended\_transits, C represents phot\_bp\_n\_obs, and D represents phot\_rp\_n\_obs. In our study, to eliminate the false positive cases where the observed source's flux is contaminated by neighboring sources, resulting in a higher BRExcess value than expected, we can select sources with a low $\beta$ value. Besides the $\beta$ value, the flux ratio between the target source and the blending source(s) should also be considered\citep{riello2021gaia}. For example, if $\beta$ is set to 0.5 and the target source has a $G_{BP}$ magnitude of 14.0 mag while the blending source has a $G_{BP}$ magnitude of 19.0 mag, the blending source's impact on the target source becomes negligible. Therefore, after selecting sources with relatively low $\beta$ value, it is crucial to assess the brightness contrast between the target and blending sources. Additionally, due to the size of {\it Gaia}'s ccd, sources influenced by nearby sourced located at angular separations larger than 1\arcsec.75 would not be classified as blended.

    In our study, we propose a screening criterion as follows: A source with a $\beta$ value $\geq$ 0.5, indicating that more than half of the observations have been classified as blended by the {\it Gaia} mission, would be categorized as significantly contaminated by neighboring sources (``abnormal"). Conversely, if the $\beta$ value is $<$ 0.5, the source is considered to be less affected by neighboring sources (``normal"). Importantly, each source that has passed through the screening process would undergo an examination of the celestial objects within a 20\arcsec.0 radius centered on the source's coordinates, to ensure the absence of contamination.
    \begin{eqnarray}
    \beta = \frac{A+B}{C+D}
    \label{eq:1}
    \end{eqnarray}
   
    \item phot\_bp(/rp)\_n\_contaminated\_transits: 
    We introduce the parameter NCT, which denotes the total number of observations classified as contaminated by the {\it Gaia}, as defined in Equation~\ref{eq:3-2} below.  
    We propose a screening criterion for NCT: a source's flux would be considered as contaminated by other sources if its NCT $>$ 0 (``abnormal"), while uncontaminated when NCT $=$ 0 (``normal).
    \begin{eqnarray}
    \label{eq:3-2}
    {\rm NCT}={\rm phot\_bp\_n\_contaminated\_transits}+\nonumber\\
{\rm phot\_rp\_n\_contaminated\_transits}
    \end{eqnarray}
\end{enumerate}

With the above 6 parameters, on one hand, we expect the HCSC candidates to exhibit ``abnormal" AEN values, indicative of extended sources. On the other hand, we anticipate the candidates to demonstrate ``normal" values across the other 5 parameters, to eliminate cases involving binaries or contaminations.

\section{Result and Discussion}\label{section:result}

Based on Figure\,\ref{fig:xfig1}, our objective was to identify potential HCSC candidates within sources with BRExcess value greater than 5 as well as those falling within the range of 2.37 to 5. The findings for sources with BRExcess $>$ 5 are detailed in Section~\ref{section:BREcess>5}, while for sources with 2.37 $<$ BRExcess $\leq$ 5 are presented in Section~\ref{section:2.37_brexcess_5}. 

\subsection{ Searching for HCSCs with BRExcess $>$ 5}\label{section:BREcess>5}
In this study, we initially tried to identify HCSCs among the 701 sources with BRExcess $>$ 5. 
We note that a number of sources are concentrated in M\,31. Given that we aims to identify HCSCs in the MW, we excluded 116 sources located in the center region of M31 (9.5° $<$ RA $<$ 12.5°, 40° $<$ Dec $<$ 42.5°).

\begin{figure*}[htbp!]
\centering
\includegraphics[scale=0.65]{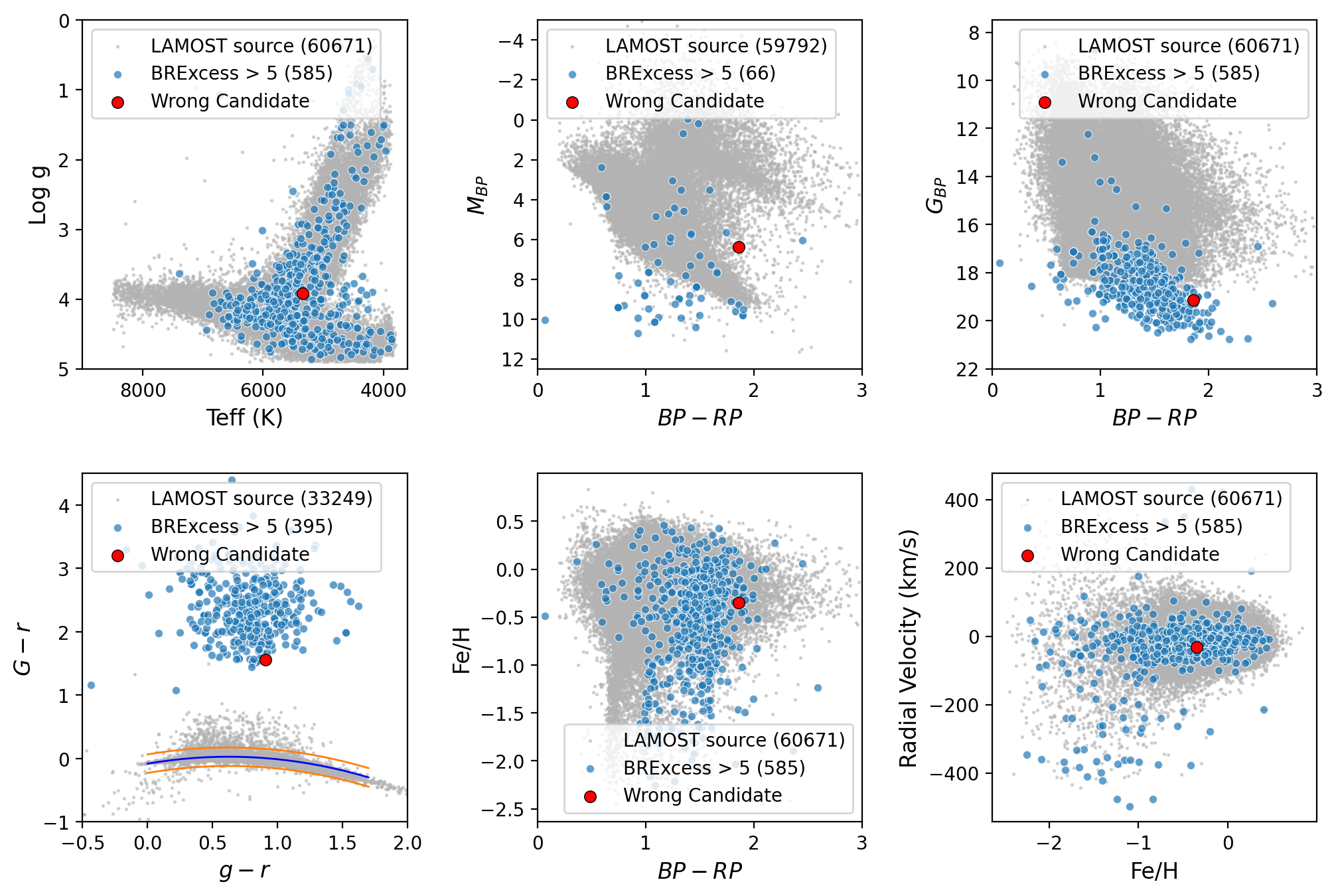}
\caption{Parameter distribution of the 585 sources with BRExcess $>$ 5, compared to a random sample of 60,671 LAMOST sources. The comparison sample is represented by gray dots, while sources with BRExcess $>$ 5, excluding those near the center of M\,31, are depicted by blue dots. The wrong candidate is indicated by the red dot. In the first subplot of the second row, the blue line is a second-order regression for $G-r$ vs $g-r$ of the comparison sample. The two orange lines indicate the position $\pm$ 3$\sigma$ away from the regression line. \label{fig:candidate_parameters_distribution}}
\end{figure*}

For the remaining 585 sources, we initially analyzed their parameter distribution in comparison to 60,671 randomly sampled LAMOST sources, as depicted in Figure\,\ref{fig:candidate_parameters_distribution}. To double-check their $G$-band photometry, we cross-matched these 585 sources and the 60671 LAMOST sources with PS1 DR1 to obtain their $g$ and $r$ band magnitudes. To minimize errors from over-brightness, we excluded sources with $g$ $<$ 14.0 mag or $r$ $<$ 14.0 mag. For the remaining sources, we plotted the $G-r$ distribution relative to $g-r$, as shown in the subplot in the second row and first column of Figure\,\ref{fig:candidate_parameters_distribution}. All the sources with BRExcess $>$ 5 lie the region 3$\sigma$ above the best fitting regression line of the comparison sample. This positioning signifies that their measured $G$-band magnitudes are fainter than they should be, aligning with our initial expectations. Therefore, in the subplot in the first row and second column, we employed the absolute magnitudes of $G_{BP}$ since we consider it to be a more reliable measure than the $G$-band magnitudes.

\begin{figure*}[htbp!]
\centering
\includegraphics[scale=0.6]{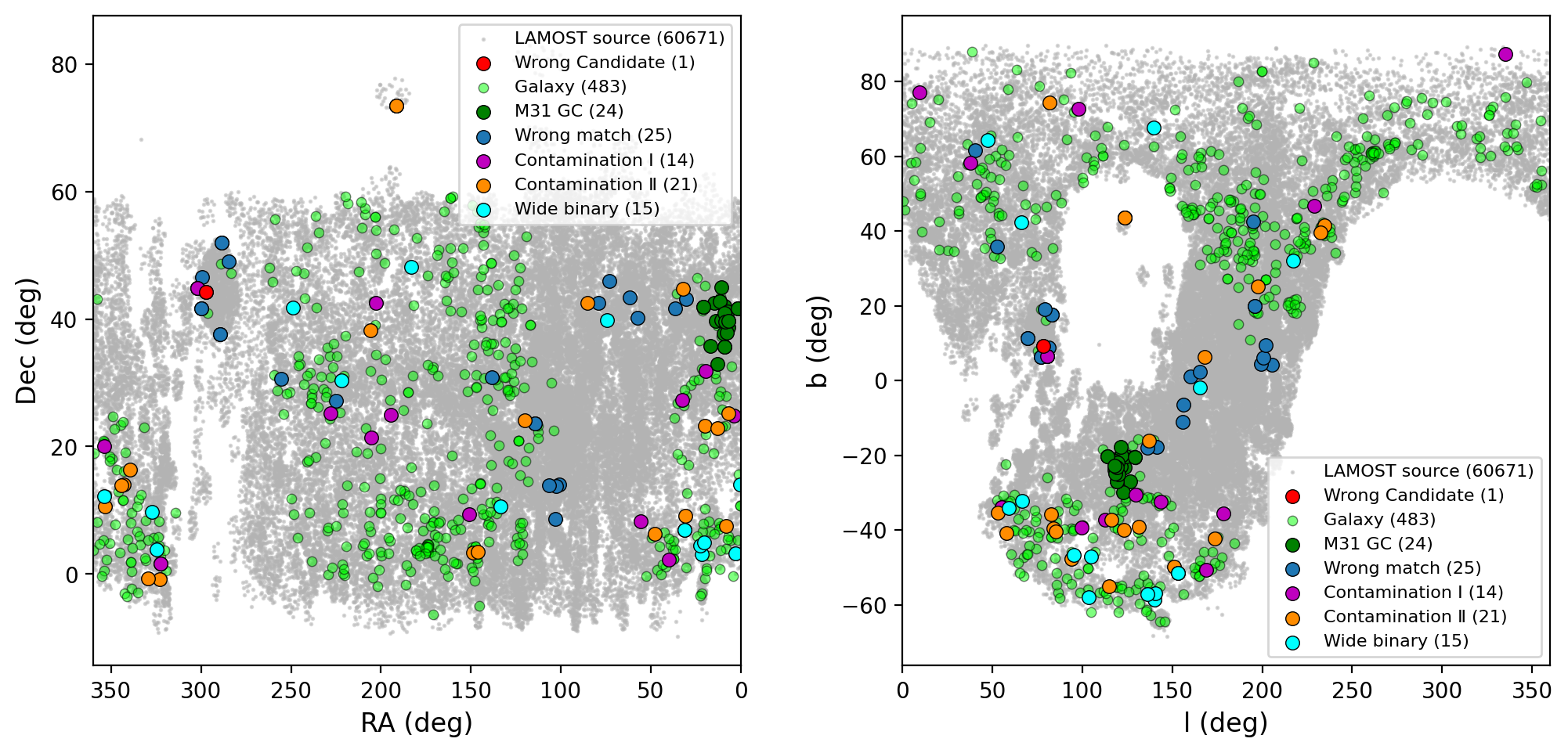}
\caption{ \label{fig:excess_5_dis}Spatial distribution of the 585 sources with BRExcess $>$ 5, with different categories identified by different colours.}
\end{figure*}

We subsequently categorized these 585 sources primarily using images and spectra from SDSS DR15. In cases where SDSS data were not accessible, we relied o PS1 images and LAMOST DR7 spectra. To facilitate our categorization, we also considered the six parameters from {\it Gaia} EDR3: AEN, RUWE, IGHA, IFMP, $\beta$, and NCT. Ultimately, we grouped these 585 sources to eight distinct categories, and their spatial distribution is shown in Figure\,\ref{fig:excess_5_dis}. 

\begin{figure*}[h!]
	\begin{center}
		\begin{minipage}{0.35\textwidth}
			\includegraphics[width=6cm,height = 6cm]{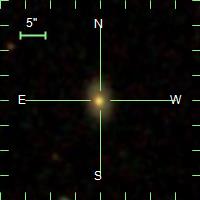}
		\end{minipage}
		\begin{minipage}{0.5\textwidth}
			\includegraphics[width=9cm,height = 6cm]{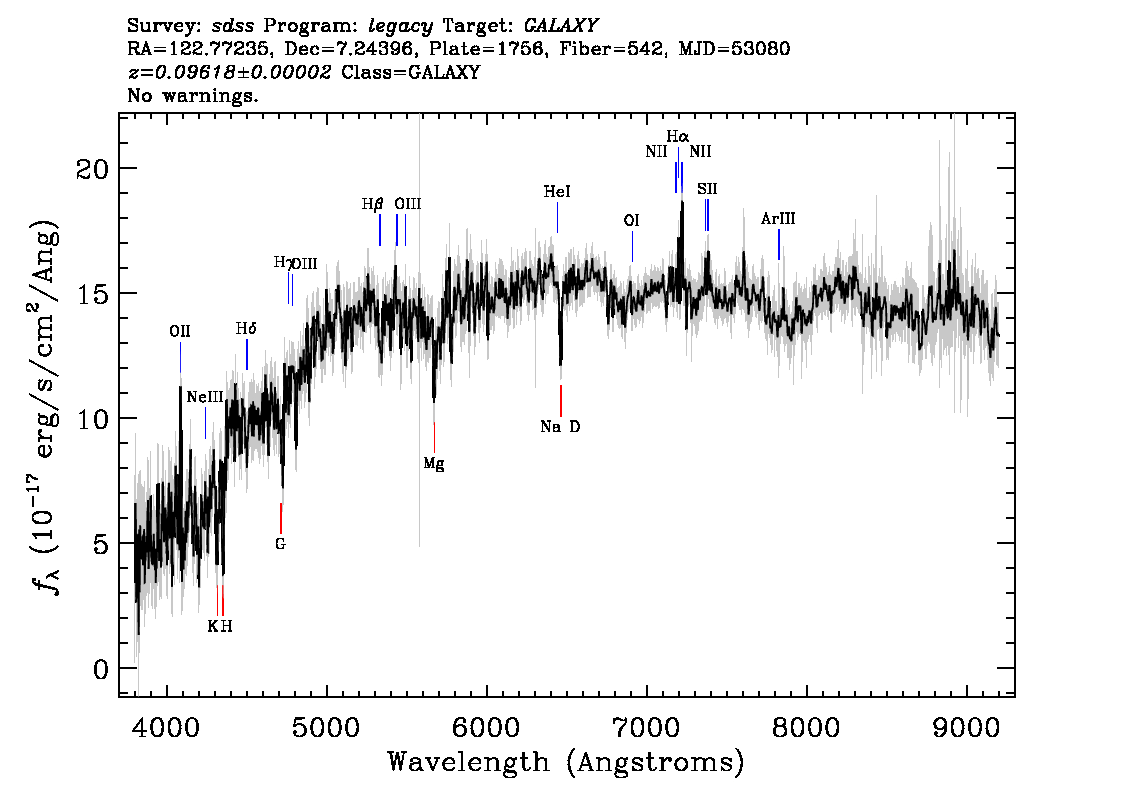}
		\end{minipage}
    \caption{An example of galaxies, whose image and spectrum are from the SDSS.}
    \label{Galaxy}
	\end{center}
\end{figure*}

Each category is outlined below:

1. Galaxies: We identified 483 sources classified as galaxies. An example is illustrated in Figure\,\ref{Galaxy}. 
Galaxies are typical kinds of extended sources, and their BRExcess values are expected to be large, as discussed in Sections~\ref{section:data_gaia} and~\ref{section:BRExcess}.

\begin{figure*}[htbp!]
	\begin{center}
		\begin{minipage}{0.35\textwidth}
			\includegraphics[width=6cm,height = 6cm]{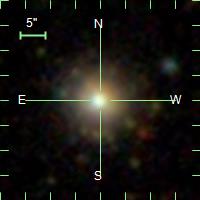}
		\end{minipage}
		\begin{minipage}{0.5\textwidth}
			\includegraphics[width=9cm,height = 6cm]{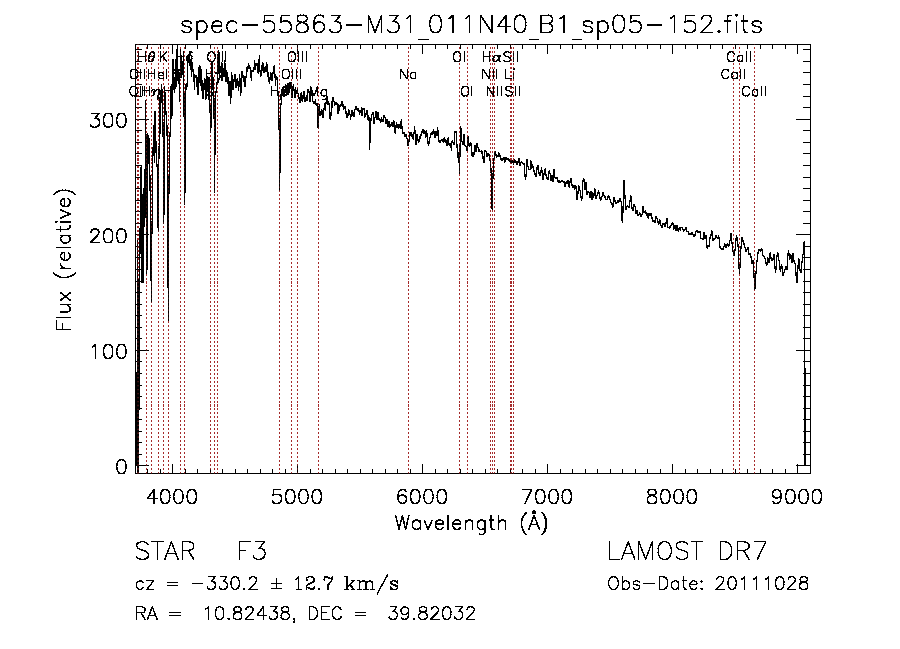}
		\end{minipage}
    \caption{An example of outer GC of M\,31, whose image is from SDSS and spectra is from LAMOST. }
    \label{M31GC}
	\end{center}
\end{figure*}

2. GCs in the outer region of M\,31 (M\,31 GCs): We have identified 24 GCs located in the outer region of M\,31. An example is shown in Figure\,\ref{M31GC}. 
The classification is based on SDSS images, LAMOST spectra, and SIMBAD information. 
GCs are also typical extended sources with large BRExcess, as described in Sections~\ref{section:data_gaia} and~\ref{section:BRExcess}.

\begin{figure*}[h]
\centering
\includegraphics[width=6cm,height = 6cm]{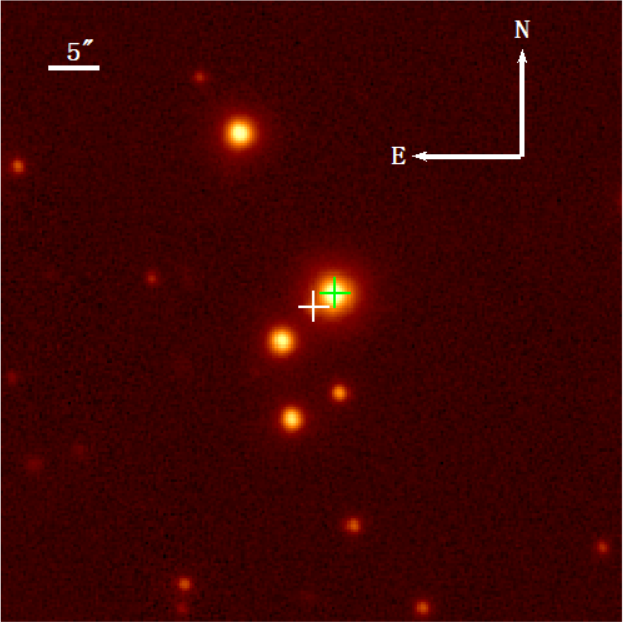}
\caption{ \label{fig:wrong_match} An example of wrong matches, whose image (1 arcmin $\times$ 1 arcmin) is from the PS1 $i$ band. The green cross denotes the location of the LAMOST source, while the white represents the {\it Gaia} source after a wrong match process.}
\end{figure*}

3. Wrong matches:
There are 25 wrong match cases, and one example is shown in Figure\,\ref{fig:wrong_match}. These cases are categorized when the magnitude of the source in the LAMOST input catalog is more than five magnitudes brighter than that of the corresponding {\it Gaia} EDR3 source after cross-matching. Combined with the images, it becomes evident that an incorrect match took place during the cross-matching process, where bright LAMOST sources were erroneously paired with faint sources of {\it Gaia} EDR3.

\begin{figure*}[htbp!]
	\begin{center}
		\begin{minipage}{0.35\textwidth}
			\includegraphics[width=6cm,height =6cm]{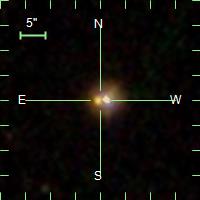}
		\end{minipage}
		\begin{minipage}{0.5\textwidth}
			\includegraphics[width=6cm,height = 6cm]{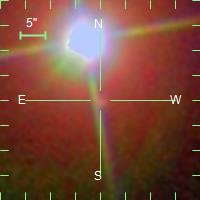}
		\end{minipage}
    \caption{ \label{fig:contamination} Left: An example of contamination \uppercase\expandafter{\romannumeral1} cases, whose image is from the SDSS. Right: An example of contamination \uppercase\expandafter{\romannumeral2} cases, whose image is from the SDSS.}
	\end{center}
\end{figure*}

4. Contamination \uppercase\expandafter{\romannumeral1}:  This category comprises 14 sources, and an example is depicted in the left panel of Figure\,\ref{fig:contamination}. These sources
exhibit characteristics such as NCT $>$ 0 or $\beta$ $>$ 0.5, indicating contamination from nearby sources. We checked the flux ratio between the source and the neighbor, confirming that the neighboring source is sufficiently bright to introduce a significantly higher flux for the target source in the $BP$ and $RP$ bands compared to its true flux. Consequently, this leads to a large BRExcess value compared to normal single stars.

5. Contamination \uppercase\expandafter{\romannumeral2}:  This category comprises 21 sources, and an example is illustrated in the right panel of Figure\,\ref{fig:contamination}. These sources exhibit NCT $=$ 0 and $\beta$ $<$ 0.1, indicating that {\it Gaia} does not detect any contaminating or blending signals. However, we identified at least one much brighter (by 5.0--11.0 mag) foreground star located within 3\arcsec.0 to 20\arcsec.0 of the target source. 

Combining the SDSS and PS1 images, we concluded that although {\it Gaia} did not detect any contamination due to the relative large angular separation, the bright foreground stars caused contamination, leading to an increased BRExcess for the target source.

\begin{figure*}[htbp!]
	\begin{center}
		\begin{minipage}{0.35\textwidth}
			\includegraphics[width=6cm,height =6cm]{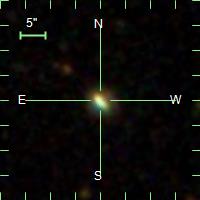}
		\end{minipage}
		\begin{minipage}{0.5\textwidth}
			\includegraphics[width=9cm,height = 6cm]{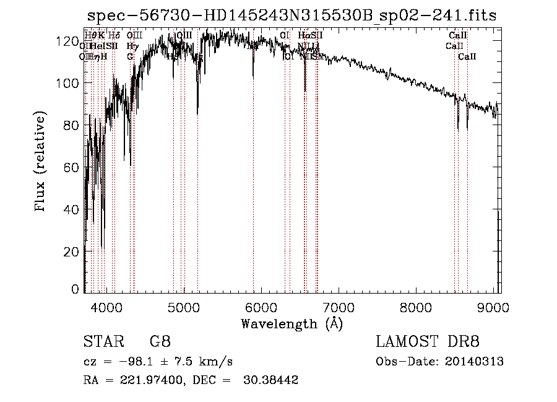}
		\end{minipage}
    \caption{An examples of wide binaries, whose image is from the SDSS and spectra is from the LAMOST. }
    \label{wide_binary}
	\end{center}
\end{figure*}

\begin{figure*}[htbp!]
\centering
\includegraphics[scale=0.75]{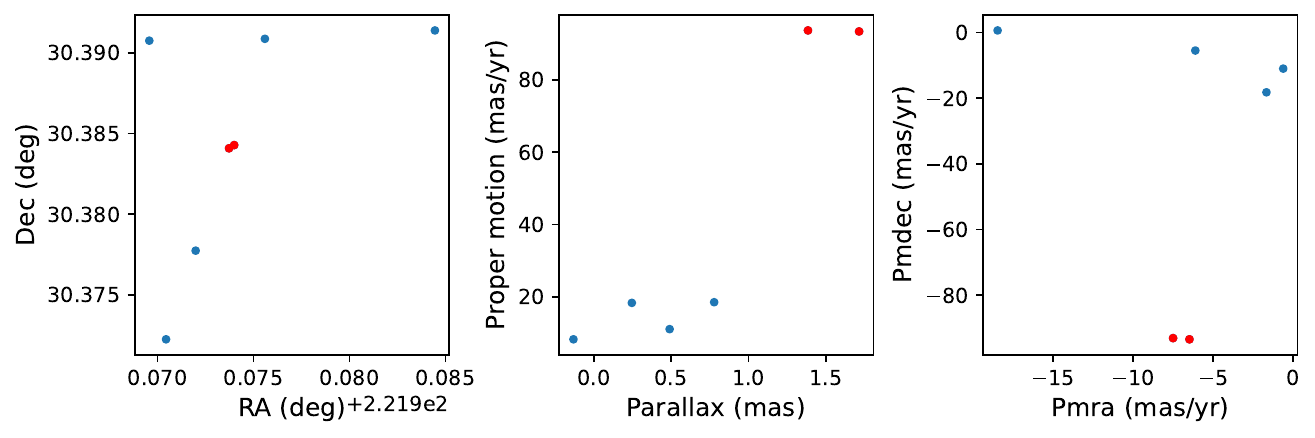}
\caption{\label{fig:widebinary_fenbu} An example of the identifications of wide binaries. Blue and red dots represent sources located within 1' of the target source in {\it Gaia} DR3. One of the two red dots is the target source, while the other is considered to be another component of a wide binary system, together with the target source. 
}
\end{figure*}

6. Wide binaries: We identified 15 wide binary systems, and an illustration of images and spectra for these systems is presented in Figure\,\ref{wide_binary}. To identify such systems, we conducted a search for nearby sources of the target source's 1' radius in the {\it Gaia} DR3 catalog, to check their spatial distributions, parallaxes and proper motions. Another source with similar positions and astrometric parameters to the target source was classified as the other component of a wide binary system. 
Due to mutual contamination between the binary components, the target source displays a large BRExcess. An illustrative example of this classification process is shown in Figure\,\ref{fig:widebinary_fenbu}.


7. Lack of data: Two in total. Due to the low image quality, we cannot classify these two sources.

\begin{figure*}[htbp!]
	\begin{center}
		\begin{minipage}{0.35\textwidth}
			\includegraphics[width=6cm,height = 6cm]{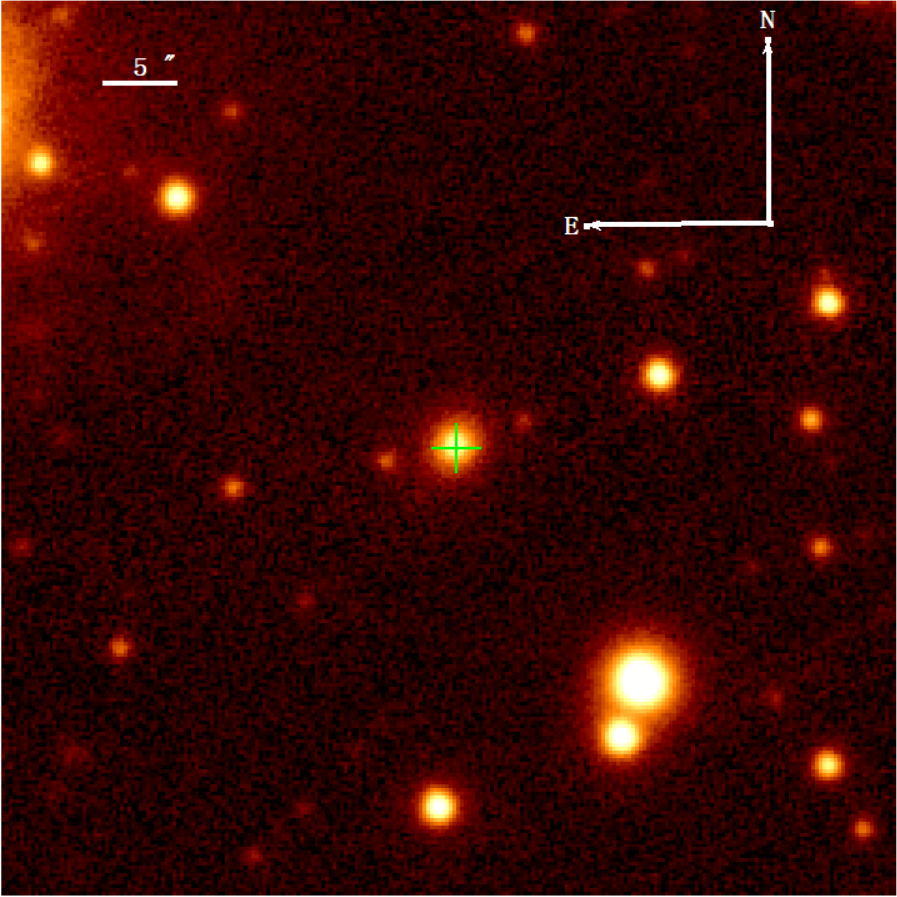}
		\end{minipage}
		\begin{minipage}{0.5\textwidth}
			\includegraphics[width=9cm,height = 6cm]{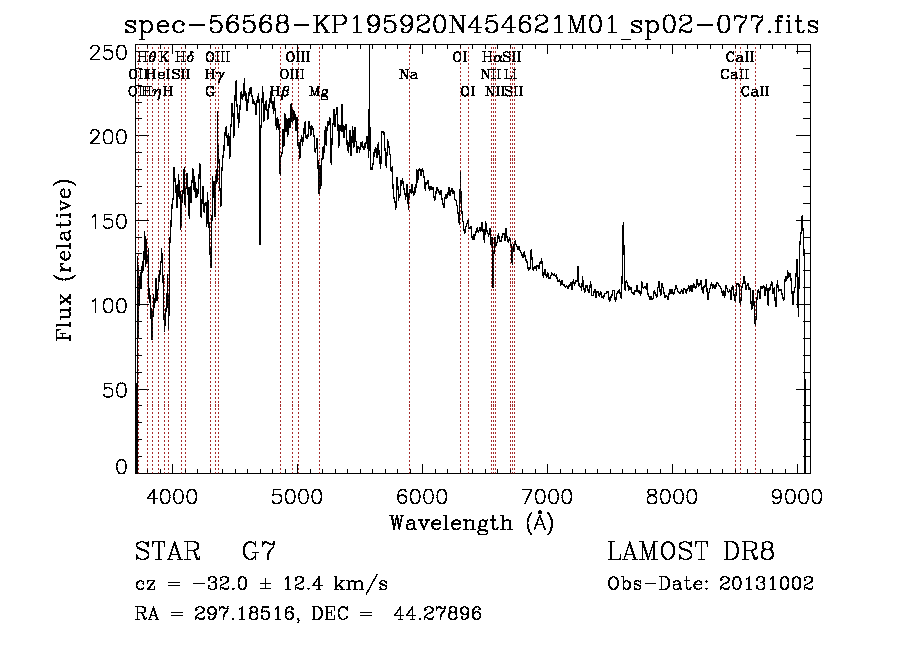}
		\end{minipage}
    \caption{Wrong candidate with BRExcess $>$ 5, whose image (1 arcmin $\times$ 1 arcmin) is from the PS1 $i$-band and the spectra is from LAMOST. The green cross in the center of the left panel denotes the candidate.}
    \label{BRExcess_5_candidate}
	\end{center}
\end{figure*}

8. Wrong candidate: Among the 585 sources, one does not fall into any of the aforementioned categories, making it stand out as a potential HCSC candidate --- an extended source with a stellar spectrum in the MW. It exhibits an AEN value greater than ${\rm AEN_{3\sigma}}$ along with an ``abnormal" RUWE value. Additionally, it exhibits ``normal" IGHA, IFMP, $\beta$ and NCT values (NCT $=$ 0 and $\beta$ $=$ 0.047). Its PS1 cutout image is shown in the left panel of Figure\,\ref{BRExcess_5_candidate}. The LAMOST spectrum identifies this source as a G7 star, as depicted on the right panel of Figure\,\ref{BRExcess_5_candidate}. 

However, we were surprised by the relatively high signal-to-noise ratio of of the LAMOST spectrum for this faint candidate. Consequently, we examined the signal-to-noise ratio in the $g$-band (${\rm S/N_{\it g}}$) as the function of the $g$ magnitude for LAMOST sources within the same plate of the candidate.  
The ${\rm S/N_{\it g}}$ value for this candidate stands out significantly, suggesting a potential anomaly. This discrepancy raises concerns about a possible acquisition error in the spectrum, indicating that the observed spectrum may not belong to this candidate.

We then examined this candidate's parameter distribution, compared to the other sources with BRExcess $>$ 5, as shown in Figure\,\ref{fig:candidate_parameters_distribution}. Given the candidate's faint absolute magnitude, we became more convinced that this is not the HCSC candidate we are searching for. Considering its image in the left panel of Figure\,\ref{BRExcess_5_candidate}, we propose that the candidate in question may actually be a galaxy and the corresponding LAMOST spectra might have been taken from the wrong target. It is essential to validate our findings by reacquiring the spectrum of this specific source.

\subsection{Searching for HCSCs with 2.37 $<$ BRExcess $\leq$ 5} \label{section:2.37_brexcess_5}
\begin{figure*}[htbp!]
\centering
\includegraphics[scale=0.65]{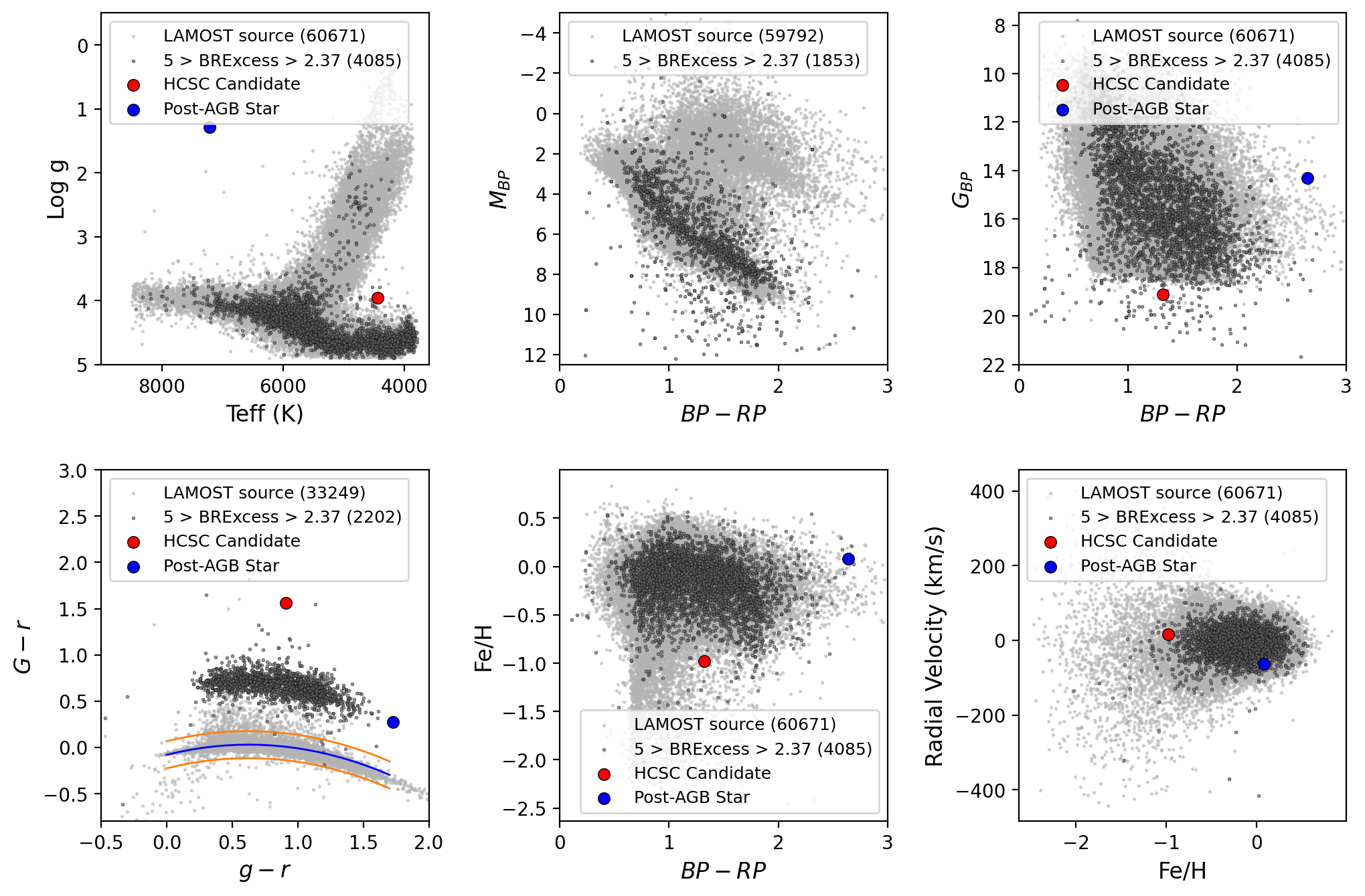}
\caption{Distribution of parameters for the candidate with 2.37 $<$ BRExcess $\leq$ 5. The light gray dots represent the 60671 random LAMOST sources, while the dark gray dots correspond to the sources with 2.37 $<$ BRExcess $\leq$ 5. The HCSC candidate is denoted as the red dot. The subplot of the first column of the second row is made similarly to that in Figure\,\ref{fig:candidate_parameters_distribution}\label{fig:candidate_parameter_1}. }
\end{figure*}

For the 4,085 sources with BRExcess values ranging from 2.37 to 5, we first analyzed their parameter distribution, following a procedure similar to that in Section~\ref{section:BREcess>5}. The results are presented in Figure\,\ref{fig:candidate_parameter_1}.

Subsequently, we employed distinct screening methods to identify single-source resolved HCSCs and multi-sources resolved HCSCs, respectively. The screening methods for each category are detailed below, in accordance with our anticipated characteristics for each type of HCSCs.

\subsubsection{ Searching for single-source resolved HCSCs}  \label{section:extremely_compact}
Firstly, in the case of the single-source resolved HCSCs, we expect {\it Gaia} to detect the entire cluster as a single extended source, without any nearby point sources. As a result, we can reliably eliminate sources with significant BRExcess resulting from neighboring sources' contamination. Our screening process depends on the {\it Gaia} EDR3 parameters listed below.

Step 1) To exclude the case of the source being part of a binary system, which could result in a large BRExcess value due to contamination from companion stars, we establish the following criteria:
\begin{enumerate}

    \item IFMP: 
    We defined two criteria, IFMP $<$ 10 as a lenient criterion and IFMP $=$ 0 as a stringent criterion.
    
    \item IGHA:  IGHA $<$ 0.2 as a lenient criterion, and IGHA $<$ 0.14 as a stringent criterion.
    
    \item RUWE: RUWE $<$ 1.25.
    
\end{enumerate}

Step 2) To minimize the possibility of targets being contaminated by nearby sources and consequently displaying large BRExcess values, we established the following criterion:
\begin{enumerate}
    \item $\beta$:  We required $\beta$ $<$ 0.5 to ensure that at least 50\% of the observations for the target are not regarded as blended by {\it Gaia}. 
    
    \item NCT: We required NCT $=$ 0. For sources with NCT $>$ 0, we considered them to be significantly contaminated from nearby sources.
\end{enumerate}
    
Step 3) From Section~\ref{section:data_gaia}, we anticipate that the AEN values of HCSCs should be 3$\sigma$ above the average AEN values of foreground stars, and therefore we implemented a screening criterion: 
\begin{enumerate}
\item AEN: ${\rm AEN_{source}}$ $>$ ${\rm AEN_{3\sigma}}$.
\end{enumerate}

Step 4) If a source satisfies all the criteria mentioned above, it undergoes a more rigorous screening process using IGHA, IFMP, and RUWE:
\begin{enumerate}
\item IGHA $<$ 0.14
\item IFMP $=$ 0 
\item RUWE $<$ 1.25
\end{enumerate}
Sources that meet all three criteria are labeled as ``3A candidates". Those meeting two of the criteria are classified as ``2A candidates". Similarly, sources meeting only one criterion are designated as ``1A candidates". Lastly, sources failing to meet any of the three criteria are denoted as ``0A candidates".

\begin{figure}[htbp!]
\centering
\includegraphics[scale=0.75]{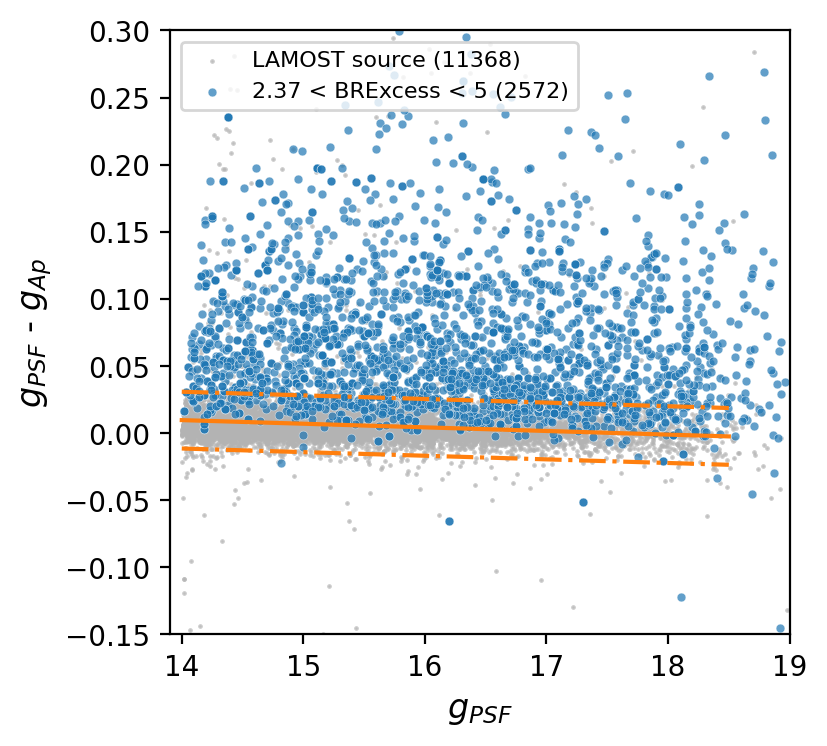}
\caption{The distribution of $\Delta$$mag_g$ with $g_{PSF}$, where grey dots represent the comparison sample with BRExcess $\leq$ 2.37, while blue dots denote sources with 2.37 $<$ BRExcess $\leq$ 5. The central solid orange line is the best linear fitting for the comparison sample, and the two dashed orange lines are $\pm$ 3$\sigma$ away from the the best fitting line.\label{fig:4085_deltagvspsf}}
\end{figure}

We also employ $PSF$ magnitudes and $Ap$ (Aperture) magnitudes of sources to make references to the characteristics as either extended or point sources. We crossmatched 4,085 sources with BRExcess values between 2.37 and 5 with PS1 DR1 to obtain their $g$-band $PSF$ and $Ap$ magnitudes. Additionally, we randomly selected 17000 LAMOST sources with BRExcess $\leq$ 2.37 as the comparison sample and obtained their $g$-band $PSF$ and $Ap$ magnitudes. 
We retained sources with magnitudes fainter than 14.0 mag to minimize errors from over-brightness. This resulted in 2,572 sources with BRExcess values between 2.37 and 5 and 11,368 sources in the comparison sample. We then explored the relationship between the magnitude difference ($\Delta mag_g$, calculated as $g_{PSF}$ - $g_{Ap}$), and $g_{PSF}$ values, as shown in Figure\,\ref{fig:4085_deltagvspsf}. 
We performed a linear regression on the $\Delta mag_g$ values of the comparison sample against $g_{PSF}$, followed by applying a Gaussian function to the residuals to determine their scatter $\sigma$. 
In Figure\,\ref{fig:4085_deltagvspsf}, the $\Delta mag_g$ values of most sources with BRExcess values between 2.37 and 5 lie above the mean of the comparison sample (represented by the best-fit) plus 3$\sigma$ (denoted as `above $+$3$\sigma$' hereafter), consistent with the expectations for extended sources. A few sources fall within the $\pm$ 3$\sigma$ range from the best-fit (denoted as `within $\pm$3$\sigma$' hereafter) or below the best-fit minus 3$\sigma$ (denoted as `below $-$3$\sigma$' hereafter), indicating a higher likelihood of being point sources in the PS1 survey. Subsequently, we applied our screening process to each of these three subsets, and the final results are presented below:
\begin{enumerate}
    \item 19 sources below $-$3$\sigma$: one 2A candidate was screened.
    \item 647 sources within $\pm$3$\sigma$: no candidate was screened.
    \item 1906 sources above $+$3$\sigma$: one 2A candidate and one 1A candidate were screened.
\end{enumerate}

Subsequently, we conducted a comprehensive validation of each candidate passed through the screening process. The validation involved the cross-referencing of various datasets, including SDSS DR15, PS1 images and LAMOST spectra. Additionally, we carefully examined sources within a 20\arcsec.0 radius centered on the candidate's coordinates in {\it Gaia} DR3 to determine if the candidate is contaminated by nearby sources. The results of this validation process are summarized in Table~\ref{tab:extremely_g}.

\setlength{\tabcolsep}{0.95mm}{
\begin{table*}[htbp!]
\footnotesize
\centering
\caption{Searching for single-source resolved HCSCs: screening results.\label{tab:extremely_g}}
\begin{tabular}{lccccccr}
\hline
Source\_id (type) & \,RA (deg)\, & \,Dec (deg)\, & \,RUWE\, & \,IGHA\, & \,IFMP (\%)\, & $\Delta$$mag_g$ & Ultimate classification \\ \hline
\multicolumn{8}{c}{Sources with $g$-band $PSF$ and $Ap$ magnitudes} \\
6498392893703424 (1A) & 41.28292 & 5.70386 & 1.5272 & 0.0538 & 3 & above $+$3$\sigma$ & HCSC candidate \\ 
2705814759231952768 (2A) & 339.53612 & 5.18785 & 1.4281 & 0.1010 & 0 below $-$3$\sigma$ & Contamination \uppercase\expandafter{\romannumeral1} \\ 
173086700992466688 (2A) & 68.23740 & 34.60345 & 5.7579 & 0.0522 & 0 & below $-$3$\sigma$ & Post-AGB Star \\ \hline
\multicolumn{8}{c}{Sources lacking $g$-band $PSF$ and $Ap$ magnitudes or over-brightness} \\
3267157265308327424 (2A) & 46.78342 & 1.08972 & 1.7812 & 0.1048 & 0 & -- & Galaxy \\ 
3848578082069151104 (1A) & 149.63648 & 3.74268 & -- & 0.1652 & 0 & -- & Contamination \uppercase\expandafter{\romannumeral2} \\ 
1484005522545507584 (1A) & 214.08919 & 36.70537 & 1.7970 & 0.1483 & 0 & -- & Galaxy \\ 
2563091316653315840 (1A) & 19.76825 & 4.54843 & 1.6664 & 0.1316 & 9 & -- & Wide binary \\ 
1565731912601035392 (1A) & 203.76656 & 57.11682 & 1.8412 & 0.0446 & 1 & -- & Galaxy \\ 
2752350592445700352 (0A) & 3.68213 & 8.87581 & 2.0088 & 0.1854 & 6 & -- & Contamination \uppercase\expandafter{\romannumeral2}\\ 
369143990894286080 (0A) & 10.41455 & 40.67588 & -- & 0.1647 & 7 & -- & M31 GC \\ 
932233479210634752 (0A) & 124.33875 & 50.88864 & 1.2872 & 0.1455 & 5 & -- & Contamination \uppercase\expandafter{\romannumeral2}\\ \hline
\end{tabular}
\end{table*}}

\begin{figure*}[htbp!]
	\begin{center}
		\begin{minipage}{0.35\textwidth}
			\includegraphics[width=6cm,height = 6cm]{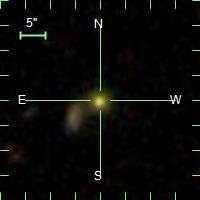}
		\end{minipage}
		\begin{minipage}{0.55\textwidth}
			\includegraphics[width=9cm,height = 6cm]{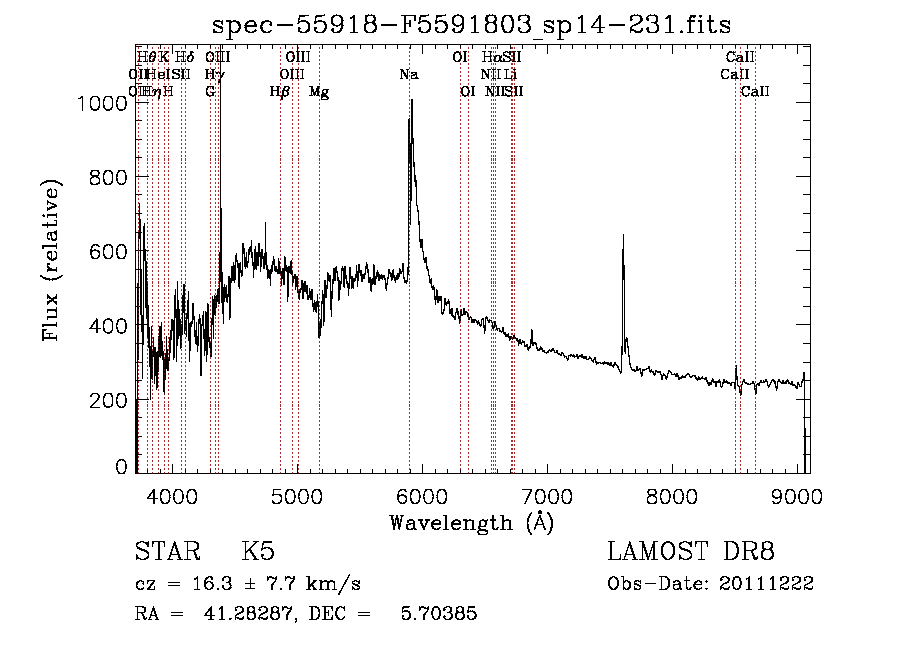}
		\end{minipage}
    \caption{Single-source resolved HCSC candidate, whose image is from SDSS and the spectra is from LAMOST. }
    \label{fig:extremely_g_candidate}
	\end{center}
\end{figure*}

Among these candidates, the source with source\_id $=$ 6498392893703424\footnote{Information from LAMOST: 1) Teff $=$ 4444.07 K; 2) Logg $=$ 3.953; 3) Fe/H $=$ $-$0.976} has been identified as a potential 1A candidate. It exhibits an ``abnormal" RUWE value, while IFMP meets lenient criteria and IGHA meets stringent criteria. Although classified as a galaxy in SDSS, LAMOST spectra identifies it as a K5 spectral type star. Therefore, we consider this source as a potential HCSC candidate. The SDSS image and LAMOST spectra of this source are presented in Figure\,\ref{fig:extremely_g_candidate}. 
We examined the distribution between ${\rm S/N_{\it g}}$ and $g$ for both this candidate and other LAMOST sources observed within the same sky region with the same plate, finding no significant anomaly. We suppose that this candidate exhibits an extended source with stellar spectra characteristics. Further high resolution spectral and image verification is necessary to confirm our findings.

\begin{figure*}[htbp!]
	\begin{center}
		\begin{minipage}{0.35\textwidth}
			\includegraphics[width=6cm,height = 6cm]{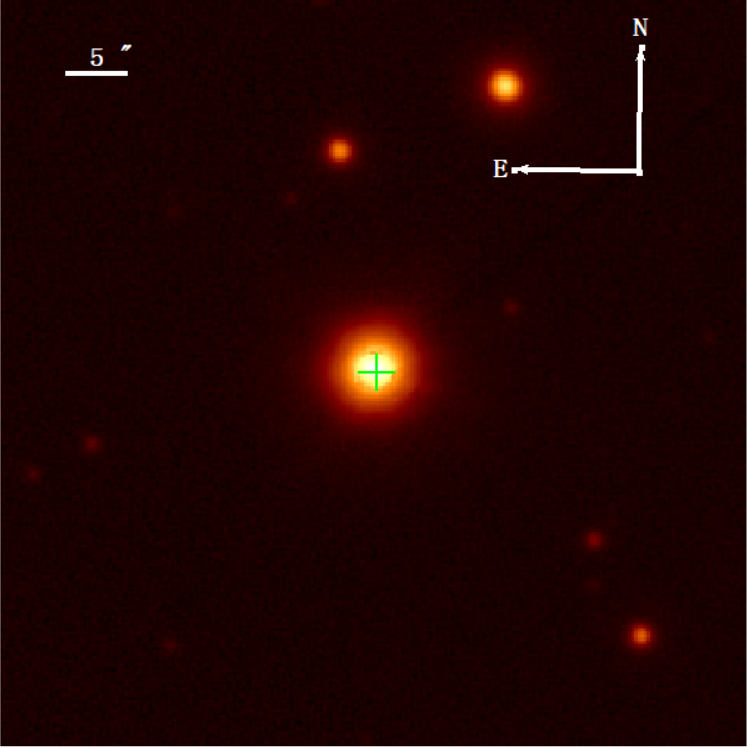}
		\end{minipage}
		\begin{minipage}{0.5\textwidth}
			\includegraphics[width=9cm,height = 6cm]{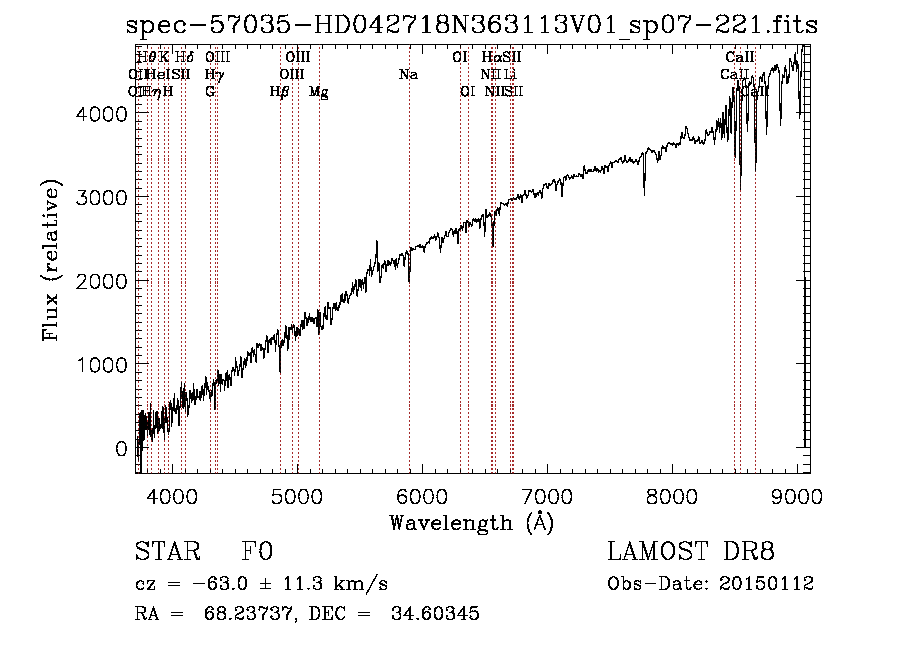}
		\end{minipage}
    \caption{Post-AGB Star: IRAS 04296+3429, whose image is from PS1 (1 arcmin $\times$ 1 arcmin) $i$-band and the spectra is from LAMOST. }
    \label{fig:PA}
	\end{center}
\end{figure*}

The source with source\_id $=$ 173086700992466688 was screened as a 2A candidate. It has been classified as a Post-AGB Star (IRAS 04296+3429) \citep{oudmaijer2022census}. We do not categorize it as a HCSC candidate, but rather as a noteworthy source deserving further investigation. In Figure\,\ref{fig:PA}, we present a cutout image centered on the source's coordinates obtained from PS1, along with the LAMOST spectra.

Subsequently, we extended the screening to the 1,511 sources initially excluded due to the absence of PS1 magnitude data or over-brightness. Within this subset, three 0A candidates, four 1A candidates, and one 2A candidate were identified. To validate these candidates, we performed a thorough validation process identical to the one mentioned earlier. The results were presented in Table~\ref{tab:extremely_g}, and no potential HCSC candidate was found in this subset.

\subsubsection{ Searching for multi-sources resolved HCSCs}  \label{section:moderately_compact}
Unlike the single-source resolved HCSCs, we expect multi-sources resolved HCSCs to display a central cluster region as a bright, extended source, accompanied by faint point sources within a 10\arcsec.0 radius. Consequently, the case of nearby sources' contamination could not be readily dismissed. Therefore, adjustments were made to the screening criteria by removing step 3). We conducted screening on three subsets categorized based on the distribution of $\Delta mag_g$ values against $g_{PSF}$, similar to Section~\ref{section:extremely_compact}, using identical notations. The outcomes are as follows, noting that candidates validated in Section~\ref{section:extremely_compact} are excluded.
\begin{enumerate}
    \item 19 sources below $-$3$\sigma$: no candidate was screened.
    \item 647 sources within $\pm$3$\sigma$: one 1A candidate was screened;
    \item 1906 sources above $+$3$\sigma$: one 2A candidate, thirteen 1A candidates, and five 0A candidates were screened. 
\end{enumerate}
We further validated each of these candidates. The results after the validation are shown in Table~\ref{tab:moderately_g}. We did not identify any HCSC candidates during this investigation. 

\setlength{\tabcolsep}{0.95mm}{
\begin{table*}[hb!]
\footnotesize
\centering
\caption{Searching for multi-sources resolved HCSCs: screening results. \label{tab:moderately_g}}
\begin{tabular}{lccccccr}
\hline
Source\_id (type) & \,RA (deg)\, & \,Dec (deg)\, & \,RUWE\, & \,IGHA\, & \,IFMP (\%)\, & $\Delta$$mag_g$ & Ultimate classification \\ \hline
\multicolumn{8}{c}{Sources with $g$-band $PSF$ and $Ap$ magnitudes} \\
449392156284056320 (2A) & 52.33586 & 57.16223 & 1.2308 & 0.0312 & 9 & above +3$\sigma$  & Wide binary \\ 
704464938231855744 (1A) & 129.78772 & 28.44897 & 1.4418 & 0.0227 & 9 & within $\pm$ 3$\sigma$ & Contamination \uppercase\expandafter{\romannumeral1} \\ 
311588474186803840 (1A) & 16.07852 & 30.45814 & 5.4640 & 0.0328 & 9 & above +3$\sigma$ & Contamination \uppercase\expandafter{\romannumeral1}\\ 
1275471219510990336 (1A) & 228.57050 & 30.12952 & 1.6151 & 0.1292 & 9 & above +3$\sigma$ & Wide binary \\ 
1528957028223827840 (1A) & 192.18811 & 44.54689 & 1.4649 & 0.0947 & 9 & above +3$\sigma$ & Wide binary \\ 
112064428424442752 (1A) & 48.24959 & 25.04831 & 1.6719 & 0.0790 & 9 & above +3$\sigma$ & Wide binary \\ 
675908353517004928 (1A) & 122.81268 & 20.80834 & 1.9797 & 0.0982 & 6 & above +3$\sigma$ & Contamination \uppercase\expandafter{\romannumeral1}\\ 
873227500510164736 (1A) & 112.04762 & 28.00048 &  -- & 0.1224 & 6 & above +3$\sigma$ & Contamination \uppercase\expandafter{\romannumeral1}\\ 
2825149467773549824 (1A) & 350.76716 & 19.77727 & 4.4414 & 0.0632 & 1 & above +3$\sigma$ & Contamination \uppercase\expandafter{\romannumeral1}\\ 
3394676631033414784 (1A) & 76.82286 & 16.23994 & 1.5533 & 0.0384 & 2 & above +3$\sigma$ & Wide binary \\ 
1572562594229422592 (1A) & 185.68549 & 54.54430 & 6.3407 & 0.0336 & 1 & above +3$\sigma$ & Wide binary \\ 
958948725548486144 (1A) & 97.43805 & 43.70542 & 1.7100 & 0.0486 & 7 & above +3$\sigma$ & Wide binary \\ 
3344212483292171520 (1A) & 92.12091 & 12.88156 & -- & 0.0717 & 6 & above +3$\sigma$ & Contamination \uppercase\expandafter{\romannumeral1}\\ 
805414227517251840 (1A) & 152.71911 & 42.19703 & 1.3181 & 0.1124 & 7 & above +3$\sigma$ & Contamination \uppercase\expandafter{\romannumeral1}\\ 
155844571964505216 (0A) & 76.51998 & 29.35847 & 1.6010 & 0.1998 & 9 & above +3$\sigma$ & Contamination \uppercase\expandafter{\romannumeral1}\\ 
883649152754683520 (0A) & 107.33776 & 27.40741 & -- & 0.1661 & 8 & above +3$\sigma$ & Contamination \uppercase\expandafter{\romannumeral1}\\ 
1586519038916853248 (0A) & 223.50317 & 44.59921 & 1.5171 & 0.1567 & 9 & above +3$\sigma$ & Wide binary \\ 
400193168469697152 (0A) & 20.03275 & 48.75375 & 1.4746 & 0.1892 & 8 & above +3$\sigma$ & Contamination \uppercase\expandafter{\romannumeral1}\\ 
1188076297256651520 (0A) & 224.16966 & 17.10362 & -- & 0.1925 & 6 & above +3$\sigma$ & Contamination \uppercase\expandafter{\romannumeral1}\\ \hline
\multicolumn{8}{c}{Sources lacking $g$-band $PSF$ and $Ap$ magnitudes or over-brightness} \\
2624183996623482624 (3A) & 342.36668 & -4.39149 & 1.1216 & 0.0545 & 0 & -- & Wide binary \\ 
141723239185156352 (2A) & 42.23868 & 37.24501 & 2.0758 & 0.0312 & 0 & -- & Wide binary \\ 
18449736397548800 (2A) & 39.63145 & 6.32936 & 1.1256 & 0.0168 & 8 & -- &Wide binary \\ 
1776969257602295040 (2A) & 333.83645 & 17.76417 & -- & 0.0462 & 0 & -- &Contamination \uppercase\expandafter{\romannumeral1}\\ 
727019735648631680 (1A) & 155.35621 & 26.09911 & 1.2713 & 0.0197 & 4 & -- &Contamination \uppercase\expandafter{\romannumeral1}\\ 
2099457980827307264 (1A) & 288.54912 & 38.56520 & 2.2538 & 0.0938 & 8 & -- &Contamination \uppercase\expandafter{\romannumeral1}\\ 
2720656997978119296 (1A) & 332.03690 & 6.64466 & 5.1655 & 0.0972 & 2 & -- &Wide binary \\ 
360649503357099392 (1A) & 10.37483 & 31.29947 & 2.6613 & 0.1983 & 0 & -- &Wide binary \\ 
640962678529014016 (1A) & 144.69539 & 21.90644 & -- & 0.0990 & 5 & -- &Contamination \uppercase\expandafter{\romannumeral1}\\ 
2763824069297410944 (1A) & 354.92190 & 11.72348 & 1.8492 & 0.0569 & 2 & -- &Wide binary \\ 
2715608178020587392 (1A) & 345.52535 & 11.32835 & -- & 0.1220 & 5 & -- &Contamination \uppercase\expandafter{\romannumeral1}\\ 
3281684081655197824 (1A) & 72.49482 & 4.89385 & 9.2288 & 0.1263 & 3 & -- &Contamination \uppercase\expandafter{\romannumeral1}\\ 
303379829611779072 (0A) & 23.45040 & 30.65799 & -- & 0.1643 & 8 & -- &Galaxy \\ 
2583656921443112576 (0A) & 18.96947 & 12.51747 & -- & 0.1564 & 7 & -- &Contamination \uppercase\expandafter{\romannumeral1}\\ \hline
\end{tabular}
\end{table*}}

We also screened the 1,511 sources lacking $g$-band magnitudes or over-brightness. Within this subset, we identified three 0A candidates, nine 1A candidates, and three 2A candidate and one 3A candidate. We conducted a comprehensive validation of these candidates. The results are detailed in Table~\ref{tab:moderately_g}, with no HCSC candidate identified.

\begin{figure*}[htbp!]
\centering
\includegraphics[scale=0.65]{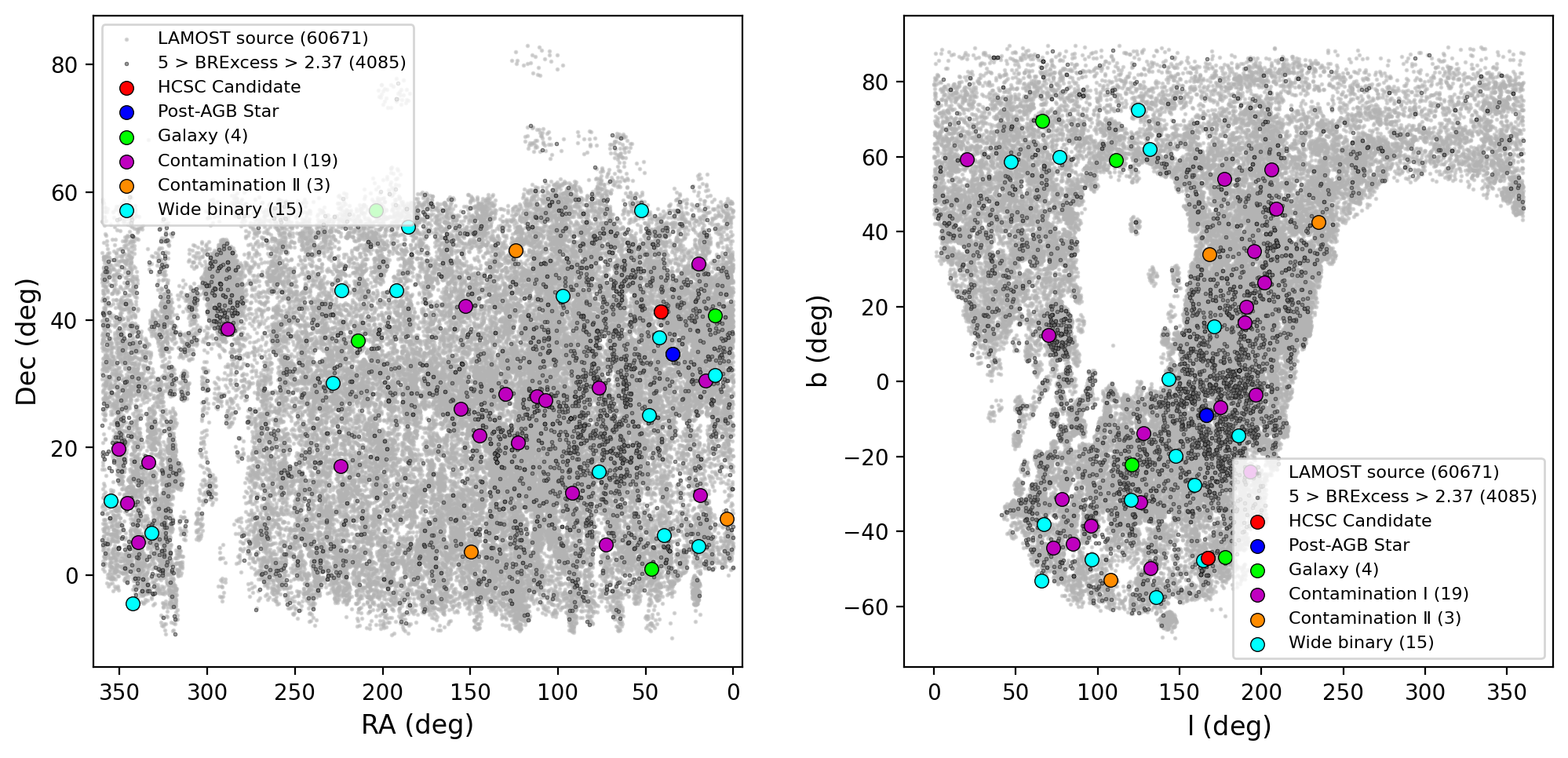}
\caption{The spatial distribution of the 4,085 sources with 2.37 $<$ BRExcess $\leq$ 5. The light gray dots represent the 60671 random LAMOST sources, while the dark gray dots correspond to the sources with 2.37 $<$ BRExcess $\leq$ 5. The HCSC candidate is represented by the red dot, while the Post-AGB star is indicated by the blue dot.\label{fig:2.37_excess_5_position} }
\end{figure*}

The spatial distribution of 4,085 sources with 2.37 $<$ BRExcess $\leq$ 5 is depicted in Figure\,\ref{fig:2.37_excess_5_position}, where all candidates passing through the screening process, including false positive cases, are differentiated by various colors. Furthermore, Figure\,\ref{fig:candidate_parameter_1} illustrates the distribution of parameters for the single-source resolved HCSC candidate and the Post-AGB star, compared to other sources with 2.37 $<$ BRExcess $\leq$ 5 and the random sampled LAMOST sources. 

To expand the identification of potential candidates, we also studied the sources with BRExcess values ranging from 1.8 to 2.37. This resulted in a dataset of 37,306 sources. Following the screening process outlined previously, we screened 6 single-source resolved HCSC candidates and 240 multi-sources resolved HCSC candidates. After conducting further verification, we found that among the 6 single-source resolved candidates, one was classified as a wide binary, two were identified as galaxies, and three were categorized to be contamination \uppercase\expandafter{\romannumeral2}. Among the 240 multi-sources resolved HCSC candidates, one was classified as a galaxy, 69 were identified as wide binary systems, and 170 were categorized as contamination \uppercase\expandafter{\romannumeral1}. We did not identify any potential HCSC candidates within this dataset. 

\section{Summary and Discussion}
\label{section:summary_discussion}
HCSCs are potential host clusters for intermediate-mass BHs. There could be hundreds of HCSCs within the MW's halo. Identifying and studying HCSCs could provide a promising way to discover and investigate intermediate-mass BHs with masses ranging from $10^4$ to $10^5$ solar masses. However, despite the exciting potential, no HCSCs have been observed in the MW to date.
 
Taking advantage of the high spatial resolution data of Gaia, we can effectively separate extended sources from point sources.  In our study, we analyzed 6,138,049 sources that are common for {\it Gaia} EDR3 and LAMOST DR7 to identify potential HCSC candidates in the MW. We further studied 701 sources with BRExcess $>$ 5 and 4,085 sources with 2.37 $<$ BRExcess $\leq$ 5.

For the 701 sources, we carried out a rigorous classification process to eliminate the false positive cases. This involved inspecting images from SDSS, PS1, examining spectra from SDSS and LAMOST, and referencing other astrometric parameters from {\it Gaia} EDR3. Among the 701 sources, the majority are galaxies, with some being M31 GCs, wide binaries, and sources contaminated by nearby objects. We identified a potential HCSC candidate, an extended source with a stellar spectrum. However, further analysis of the signal-to-noise ratio of its LAMOST spectrum led us to deduce that the spectrum does not belong to this candidate. Taking this information into account, along with its faint absolute magnitude and image characteristics, we determined it not to be an HCSC candidate. Reacquiring the source's spectrum could validate our findings.

For the 4,085 sources with BRExcess values ranging between 2.37 and 5, we tried to identify single-source resolved and multi-sources resolved HCSC candidates, respectively. We employed different {\it Gaia} EDR3 astrometric parameters to identify these two types of candidates separately. Following the identification, we conducted further validation of the candidates to eliminate any false positive cases. This validation process incorporated data from SDSS, PS1 images, LAMOST spectra, and examination of sources within a 20\arcsec.0 radius centered on each candidate's coordinates in {\it Gaia} DR3. As a result, we identified one single-source resolved HCSC candidate, which is an extended source with a star-like spectrum. Additionally, we identified a Post-AGB Star as a noteworthy source for further investigation.

The HCSC candidates can be further validated through the following observations:
\begin{enumerate}
    \item High-resolution imaging observations. We can employ instruments like Hubble Space Telescope, the Chinese Space Station Telescope (CSST; \citealt{Zhan2011}), and other high-resolution space telescopes. It will allow us to thoroughly examine the candidates, confirming their identities through distinct morphological features and structural characteristics.
    \item High-resolution spectroscopic observations. HCSCs are distinguished by high velocity dispersion. It is expected that the spectral lines in their spectra should exhibit broader profiles compared to those observed in typical stellar spectra. This spectroscopic analysis will serve as an essential criterion for certifying the HCSC candidates, providing insights into their internal dynamics.
\end{enumerate} 

We have developed a comprehensive selection procedure for identifying HCSC candidates, offering valuable insights for future data-mining efforts in upcoming data releases and surveys such as the CSST, the Nancy Grace Roman Space
Telescope \citep{Spergel2015}, and Euclid \citep{laureijs2011euclid}. Additionally, our methodology can aid in the development of search strategies for similar objects. 

\section*{acknowledgments}

This work is supported by the National Natural Science Foundation of China through the project NSFC 12222301, 12173007, 12090040, 12090044 and 11603002, and the National Key R\&D Program of China via 2019YFA0405500 and 2019YFA0405503.
We acknowledge the science research grants from the China Manned Space Project with NO. CMS-CSST-2021-A08 and CMS-CSST-2021-A09.

This work has made use of data from the European Space Agency (ESA) mission {\it Gaia} (\url{https://www.cosmos.esa.int/gaia}), processed by the {\it Gaia} Data Processing and Analysis Consortium (DPAC, \url{https:// www.cosmos.esa.int/web/gaia/dpac/ consortium}). Funding for the DPAC has been provided by national institutions, in particular the institutions participating in the {\it Gaia} Multilateral Agreement. 

This work has made use of data products from the Guoshoujing Telescope (the Large Sky Area Multi-Object Fiber Spectroscopic Telescope, LAMOST). LAMOST is a National Major Scientific Project built by the Chinese Academy of Sciences. Funding for the project has been provided by the National Development and Reform Commission. LAMOST is operated and managed by the National Astronomical Observatories, Chinese Academy of Sciences.

\bibliography{wh2023a}

\begin{thebibliography}{}
\expandafter\ifx\csname natexlab\endcsname\relax\def\natexlab#1{#1}\fi
\providecommand{\url}[1]{\href{#1}{#1}}
\providecommand{\dodoi}[1]{doi:~\href{http://doi.org/#1}{\nolinkurl{#1}}}
\providecommand{\doeprint}[1]{\href{http://ascl.net/#1}{\nolinkurl{http://ascl.net/#1}}}
\providecommand{\doarXiv}[1]{\href{https://arxiv.org/abs/#1}{\nolinkurl{https://arxiv.org/abs/#1}}}

\bibitem[{{Abazajian} {et~al.}(2009){Abazajian}, {Adelman-McCarthy},
  {Ag{\"u}eros}, {Allam}, {Allende Prieto}, {An}, {Anderson}, {Anderson},
  {Annis}, {Bahcall}, {Bailer-Jones}, {Barentine}, {Bassett}, {Becker},
  {Beers}, {Bell}, {Belokurov}, {Berlind}, {Berman}, {Bernardi}, {Bickerton},
  {Bizyaev}, {Blakeslee}, {Blanton}, {Bochanski}, {Boroski}, {Brewington},
  {Brinchmann}, {Brinkmann}, {Brunner}, {Budav{\'a}ri}, {Carey}, {Carliles},
  {Carr}, {Castander}, {Cinabro}, {Connolly}, {Csabai}, {Cunha}, {Czarapata},
  {Davenport}, {de Haas}, {Dilday}, {Doi}, {Eisenstein}, {Evans}, {Evans},
  {Fan}, {Friedman}, {Frieman}, {Fukugita}, {G{\"a}nsicke}, {Gates},
  {Gillespie}, {Gilmore}, {Gonzalez}, {Gonzalez}, {Grebel}, {Gunn},
  {Gy{\"o}ry}, {Hall}, {Harding}, {Harris}, {Harvanek}, {Hawley}, {Hayes},
  {Heckman}, {Hendry}, {Hennessy}, {Hindsley}, {Hoblitt}, {Hogan}, {Hogg},
  {Holtzman}, {Hyde}, {Ichikawa}, {Ichikawa}, {Im}, {Ivezi{\'c}}, {Jester},
  {Jiang}, {Johnson}, {Jorgensen}, {Juri{\'c}}, {Kent}, {Kessler}, {Kleinman},
  {Knapp}, {Konishi}, {Kron}, {Krzesinski}, {Kuropatkin}, {Lampeitl},
  {Lebedeva}, {Lee}, {Lee}, {French Leger}, {L{\'e}pine}, {Li}, {Lima}, {Lin},
  {Long}, {Loomis}, {Loveday}, {Lupton}, {Magnier}, {Malanushenko},
  {Malanushenko}, {Mandelbaum}, {Margon}, {Marriner}, {Mart{\'\i}nez-Delgado},
  {Matsubara}, {McGehee}, {McKay}, {Meiksin}, {Morrison}, {Mullally}, {Munn},
  {Murphy}, {Nash}, {Nebot}, {Neilsen}, {Newberg}, {Newman}, {Nichol},
  {Nicinski}, {Nieto-Santisteban}, {Nitta}, {Okamura}, {Oravetz}, {Ostriker},
  {Owen}, {Padmanabhan}, {Pan}, {Park}, {Pauls}, {Peoples}, {Percival}, {Pier},
  {Pope}, {Pourbaix}, {Price}, {Purger}, {Quinn}, {Raddick}, {Re Fiorentin},
  {Richards}, {Richmond}, {Riess}, {Rix}, {Rockosi}, {Sako}, {Schlegel},
  {Schneider}, {Scholz}, {Schreiber}, {Schwope}, {Seljak}, {Sesar}, {Sheldon},
  {Shimasaku}, {Sibley}, {Simmons}, {Sivarani}, {Allyn Smith}, {Smith},
  {Smol{\v{c}}i{\'c}}, {Snedden}, {Stebbins}, {Steinmetz}, {Stoughton},
  {Strauss}, {SubbaRao}, {Suto}, {Szalay}, {Szapudi}, {Szkody}, {Tanaka},
  {Tegmark}, {Teodoro}, {Thakar}, {Tremonti}, {Tucker}, {Uomoto}, {Vanden
  Berk}, {Vandenberg}, {Vidrih}, {Vogeley}, {Voges}, {Vogt}, {Wadadekar},
  {Watters}, {Weinberg}, {West}, {White}, {Wilhite}, {Wonders}, {Yanny},
  {Yocum}, {York}, {Zehavi}, {Zibetti}, \& {Zucker}}]{Abazajian2009}
{Abazajian}, K.~N., {Adelman-McCarthy}, J.~K., {Ag{\"u}eros}, M.~A., {et~al.}
  2009, \apjs, 182, 543, \dodoi{10.1088/0067-0049/182/2/543}

\bibitem[{{Aguado} {et~al.}(2019){Aguado}, {Ahumada}, {Almeida}, {Anderson},
  {Andrews}, {Anguiano}, {Aquino Ort{\'\i}z}, {Arag{\'o}n-Salamanca},
  {Argudo-Fern{\'a}ndez}, {Aubert}, {Avila-Reese}, {Badenes}, {Barboza
  Rembold}, {Barger}, {Barrera-Ballesteros}, {Bates}, {Bautista}, {Beaton},
  {Beers}, {Belfiore}, {Bernardi}, {Bershady}, {Beutler}, {Bird}, {Bizyaev},
  {Blanc}, {Blanton}, {Blomqvist}, {Bolton}, {Boquien}, {Borissova}, {Bovy},
  {Brandt}, {Brinkmann}, {Brownstein}, {Bundy}, {Burgasser}, {Byler}, {Cano
  Diaz}, {Cappellari}, {Carrera}, {Cervantes Sodi}, {Chen}, {Cherinka}, {Choi},
  {Chung}, {Coffey}, {Comerford}, {Comparat}, {Covey}, {da Silva Ilha}, {da
  Costa}, {Dai}, {Damke}, {Darling}, {Davies}, {Dawson}, {de Sainte Agathe},
  {Deconto Machado}, {Del Moro}, {De Lee}, {Diamond-Stanic}, {Dom{\'\i}nguez
  S{\'a}nchez}, {Donor}, {Drory}, {du Mas des Bourboux}, {Duckworth}, {Dwelly},
  {Ebelke}, {Emsellem}, {Escoffier}, {Fern{\'a}ndez-Trincado}, {Feuillet},
  {Fischer}, {Fleming}, {Fraser-McKelvie}, {Freischlad}, {Frinchaboy}, {Fu},
  {Galbany}, {Garcia-Dias}, {Garc{\'\i}a-Hern{\'a}ndez}, {Garma Oehmichen},
  {Geimba Maia}, {Gil-Mar{\'\i}n}, {Grabowski}, {Gu}, {Guo}, {Ha},
  {Harrington}, {Hasselquist}, {Hayes}, {Hearty}, {Hernandez Toledo}, {Hicks},
  {Hogg}, {Holley-Bockelmann}, {Holtzman}, {Hsieh}, {Hunt}, {Hwang},
  {Ibarra-Medel}, {Jimenez Angel}, {Johnson}, {Jones}, {J{\"o}nsson},
  {Kinemuchi}, {Kollmeier}, {Krawczyk}, {Kreckel}, {Kruk}, {Lacerna}, {Lan},
  {Lane}, {Law}, {Lee}, {Li}, {Lian}, {Lin}, {Lin}, {Lintott}, {Long},
  {Longa-Pe{\~n}a}, {Mackereth}, {de la Macorra}, {Majewski}, {Malanushenko},
  {Manchado}, {Maraston}, {Mariappan}, {Marinelli}, {Marques-Chaves},
  {Masseron}, {Masters}, {McDermid}, {Medina Pe{\~n}a}, {Meneses-Goytia},
  {Merloni}, {Merrifield}, {Meszaros}, {Minniti}, {Minsley}, {Muna}, {Myers},
  {Nair}, {Correa do Nascimento}, {Newman}, {Nitschelm}, {Olmstead}, {Oravetz},
  {Oravetz}, {Ortega Minakata}, {Pace}, {Padilla}, {Palicio}, {Pan}, {Pan},
  {Parikh}, {Parker}, {Peirani}, {Penny}, {Percival}, {Perez-Fournon},
  {Peterken}, {Pinsonneault}, {Prakash}, {Raddick}, {Raichoor}, {Riffel},
  {Riffel}, {Rix}, {Robin}, {Roman-Lopes}, {Rose}, {Ross}, {Rossi}, {Rowlands},
  {Rubin}, {S{\'a}nchez}, {S{\'a}nchez-Gallego}, {Sayres}, {Schaefer},
  {Schiavon}, {Schimoia}, {Schlafly}, {Schlegel}, {Schneider}, {Schultheis},
  {Seo}, {Shamsi}, {Shao}, {Shen}, {Shetty}, {Simonian}, {Smethurst}, {Sobeck},
  {Souter}, {Spindler}, {Stark}, {Stassun}, {Steinmetz}, {Storchi-Bergmann},
  {Stringfellow}, {Su{\'a}rez}, {Sun}, {Taghizadeh-Popp}, {Talbot}, {Tayar},
  {Thakar}, {Thomas}, {Tissera}, {Tojeiro}, {Troup}, {Unda-Sanzana},
  {Valenzuela}, {Vargas-Maga{\~n}a}, {V{\'a}zquez-Mata}, {Wake}, {Weaver},
  {Weijmans}, {Westfall}, {Wild}, {Wilson}, {Woods}, {Yan}, {Yang}, {Zamora},
  {Zasowski}, {Zhang}, {Zheng}, {Zheng}, {Zhu}, {Zinn}, \&
  {Zou}}]{aguado2019fifteenth}
{Aguado}, D.~S., {Ahumada}, R., {Almeida}, A., {et~al.} 2019, \apjs, 240, 23,
  \dodoi{10.3847/1538-4365/aaf651}

\bibitem[{{Andrew} {et~al.}(2022){Andrew}, {Penoyre}, {Belokurov}, {Evans}, \&
  {Oh}}]{Andrew2022}
{Andrew}, S., {Penoyre}, Z., {Belokurov}, V., {Evans}, N.~W., \& {Oh}, S. 2022,
  \mnras, 516, 3661, \dodoi{10.1093/mnras/stac2532}

\bibitem[{{Bekenstein}(1973)}]{Bekenstein1973}
{Bekenstein}, J.~D. 1973, \apj, 183, 657, \dodoi{10.1086/152255}

\bibitem[{{Campanelli} {et~al.}(2007){Campanelli}, {Lousto}, {Zlochower}, \&
  {Merritt}}]{Campanelli2007}
{Campanelli}, M., {Lousto}, C., {Zlochower}, Y., \& {Merritt}, D. 2007, \apjl,
  659, L5, \dodoi{10.1086/516712}

\bibitem[{{Chambers} {et~al.}(2016){Chambers}, {Magnier}, {Metcalfe},
  {Flewelling}, {Huber}, {Waters}, {Denneau}, {Draper}, {Farrow}, {Finkbeiner},
  {Holmberg}, {Koppenhoefer}, {Price}, {Rest}, {Saglia}, {Schlafly}, {Smartt},
  {Sweeney}, {Wainscoat}, {Burgett}, {Chastel}, {Grav}, {Heasley}, {Hodapp},
  {Jedicke}, {Kaiser}, {Kudritzki}, {Luppino}, {Lupton}, {Monet}, {Morgan},
  {Onaka}, {Shiao}, {Stubbs}, {Tonry}, {White}, {Ba{\~n}ados}, {Bell},
  {Bender}, {Bernard}, {Boegner}, {Boffi}, {Botticella}, {Calamida},
  {Casertano}, {Chen}, {Chen}, {Cole}, {Deacon}, {Frenk}, {Fitzsimmons},
  {Gezari}, {Gibbs}, {Goessl}, {Goggia}, {Gourgue}, {Goldman}, {Grant},
  {Grebel}, {Hambly}, {Hasinger}, {Heavens}, {Heckman}, {Henderson}, {Henning},
  {Holman}, {Hopp}, {Ip}, {Isani}, {Jackson}, {Keyes}, {Koekemoer}, {Kotak},
  {Le}, {Liska}, {Long}, {Lucey}, {Liu}, {Martin}, {Masci}, {McLean}, {Mindel},
  {Misra}, {Morganson}, {Murphy}, {Obaika}, {Narayan}, {Nieto-Santisteban},
  {Norberg}, {Peacock}, {Pier}, {Postman}, {Primak}, {Rae}, {Rai}, {Riess},
  {Riffeser}, {Rix}, {R{\"o}ser}, {Russel}, {Rutz}, {Schilbach}, {Schultz},
  {Scolnic}, {Strolger}, {Szalay}, {Seitz}, {Small}, {Smith}, {Soderblom},
  {Taylor}, {Thomson}, {Taylor}, {Thakar}, {Thiel}, {Thilker}, {Unger},
  {Urata}, {Valenti}, {Wagner}, {Walder}, {Walter}, {Watters}, {Werner},
  {Wood-Vasey}, \& {Wyse}}]{chambers2016pan}
{Chambers}, K.~C., {Magnier}, E.~A., {Metcalfe}, N., {et~al.} 2016, arXiv
  e-prints, arXiv:1612.05560, \dodoi{10.48550/arXiv.1612.05560}

\bibitem[{{Cui} {et~al.}(2012){Cui}, {Zhao}, {Chu}, {Li}, {Li}, {Zhang}, {Su},
  {Yao}, {Wang}, {Xing}, {Li}, {Zhu}, {Wang}, {Gu}, {Luo}, {Xu}, {Zhang},
  {Liu}, {Zhang}, {Yang}, {Cao}, {Chen}, {Chen}, {Chen}, {Chen}, {Chu}, {Feng},
  {Gong}, {Hou}, {Hu}, {Hu}, {Hu}, {Jia}, {Jiang}, {Jiang}, {Jiang}, {Jin},
  {Li}, {Li}, {Li}, {Liu}, {Liu}, {Lu}, {Mao}, {Men}, {Qi}, {Qi}, {Shi},
  {Tang}, {Tao}, {Wang}, {Wang}, {Wang}, {Wang}, {Wang}, {Wang}, {Wang},
  {Wang}, {Wang}, {Wang}, {Wang}, {Wang}, {Xu}, {Xu}, {Yang}, {Yu}, {Yuan},
  {Yuan}, {Zhai}, {Zhang}, {Zhang}, {Zhang}, {Zhao}, {Zhou}, {Zhou}, {Zhu}, \&
  {Zou}}]{cui2012large}
{Cui}, X.-Q., {Zhao}, Y.-H., {Chu}, Y.-Q., {et~al.} 2012, Research in Astronomy
  and Astrophysics, 12, 1197, \dodoi{10.1088/1674-4527/12/9/003}

\bibitem[{{De Angeli} {et~al.}(2023){De Angeli}, {Weiler}, {Montegriffo},
  {Evans}, {Riello}, {Andrae}, {Carrasco}, {Busso}, {Burgess}, {Cacciari},
  {Davidson}, {Harrison}, {Hodgkin}, {Jordi}, {Osborne}, {Pancino},
  {Altavilla}, {Barstow}, {Bailer-Jones}, {Bellazzini}, {Brown}, {Castellani},
  {Cowell}, {Delchambre}, {De Luise}, {Diener}, {Fabricius}, {Fouesneau},
  {Fr{\'e}mat}, {Gilmore}, {Giuffrida}, {Hambly}, {Hidalgo}, {Holland},
  {Kostrzewa-Rutkowska}, {van Leeuwen}, {Lobel}, {Marinoni}, {Miller},
  {Pagani}, {Palaversa}, {Piersimoni}, {Pulone}, {Ragaini}, {Rainer},
  {Richards}, {Rixon}, {Ruz-Mieres}, {Sanna}, {Sarro}, {Rowell}, {Sordo},
  {Walton}, \& {Yoldas}}]{de2022gaia}
{De Angeli}, F., {Weiler}, M., {Montegriffo}, P., {et~al.} 2023, \aap, 674, A2,
  \dodoi{10.1051/0004-6361/202243680}

\bibitem[{{Dey} {et~al.}(2019){Dey}, {Schlegel}, {Lang}, {Blum}, {Burleigh},
  {Fan}, {Findlay}, {Finkbeiner}, {Herrera}, {Juneau}, {Landriau}, {Levi},
  {McGreer}, {Meisner}, {Myers}, {Moustakas}, {Nugent}, {Patej}, {Schlafly},
  {Walker}, {Valdes}, {Weaver}, {Y{\`e}che}, {Zou}, {Zhou}, {Abareshi},
  {Abbott}, {Abolfathi}, {Aguilera}, {Alam}, {Allen}, {Alvarez}, {Annis},
  {Ansarinejad}, {Aubert}, {Beechert}, {Bell}, {BenZvi}, {Beutler}, {Bielby},
  {Bolton}, {Brice{\~n}o}, {Buckley-Geer}, {Butler}, {Calamida}, {Carlberg},
  {Carter}, {Casas}, {Castander}, {Choi}, {Comparat}, {Cukanovaite}, {Delubac},
  {DeVries}, {Dey}, {Dhungana}, {Dickinson}, {Ding}, {Donaldson}, {Duan},
  {Duckworth}, {Eftekharzadeh}, {Eisenstein}, {Etourneau}, {Fagrelius},
  {Farihi}, {Fitzpatrick}, {Font-Ribera}, {Fulmer}, {G{\"a}nsicke},
  {Gaztanaga}, {George}, {Gerdes}, {Gontcho}, {Gorgoni}, {Green}, {Guy},
  {Harmer}, {Hernandez}, {Honscheid}, {Huang}, {James}, {Jannuzi}, {Jiang},
  {Joyce}, {Karcher}, {Karkar}, {Kehoe}, {Kneib}, {Kueter-Young}, {Lan},
  {Lauer}, {Le Guillou}, {Le Van Suu}, {Lee}, {Lesser}, {Perreault Levasseur},
  {Li}, {Mann}, {Marshall}, {Mart{\'\i}nez-V{\'a}zquez}, {Martini}, {du Mas des
  Bourboux}, {McManus}, {Meier}, {M{\'e}nard}, {Metcalfe},
  {Mu{\~n}oz-Guti{\'e}rrez}, {Najita}, {Napier}, {Narayan}, {Newman}, {Nie},
  {Nord}, {Norman}, {Olsen}, {Paat}, {Palanque-Delabrouille}, {Peng},
  {Poppett}, {Poremba}, {Prakash}, {Rabinowitz}, {Raichoor}, {Rezaie},
  {Robertson}, {Roe}, {Ross}, {Ross}, {Rudnick}, {Safonova}, {Saha},
  {S{\'a}nchez}, {Savary}, {Schweiker}, {Scott}, {Seo}, {Shan}, {Silva},
  {Slepian}, {Soto}, {Sprayberry}, {Staten}, {Stillman}, {Stupak}, {Summers},
  {Sien Tie}, {Tirado}, {Vargas-Maga{\~n}a}, {Vivas}, {Wechsler}, {Williams},
  {Yang}, {Yang}, {Yapici}, {Zaritsky}, {Zenteno}, {Zhang}, {Zhang}, {Zhou}, \&
  {Zhou}}]{dey2019overview}
{Dey}, A., {Schlegel}, D.~J., {Lang}, D., {et~al.} 2019, \aj, 157, 168,
  \dodoi{10.3847/1538-3881/ab089d}

\bibitem[{{Eisenstein} {et~al.}(2011){Eisenstein}, {Weinberg}, {Agol},
  {Aihara}, {Allende Prieto}, {Anderson}, {Arns}, {Aubourg}, {Bailey},
  {Balbinot}, {Barkhouser}, {Beers}, {Berlind}, {Bickerton}, {Bizyaev},
  {Blanton}, {Bochanski}, {Bolton}, {Bosman}, {Bovy}, {Brandt}, {Breslauer},
  {Brewington}, {Brinkmann}, {Brown}, {Brownstein}, {Burger}, {Busca},
  {Campbell}, {Cargile}, {Carithers}, {Carlberg}, {Carr}, {Chang}, {Chen},
  {Chiappini}, {Comparat}, {Connolly}, {Cortes}, {Croft}, {Cunha}, {da Costa},
  {Davenport}, {Dawson}, {De Lee}, {Porto de Mello}, {de Simoni}, {Dean},
  {Dhital}, {Ealet}, {Ebelke}, {Edmondson}, {Eiting}, {Escoffier}, {Esposito},
  {Evans}, {Fan}, {Femen{\'\i}a Castell{\'a}}, {Dutra Ferreira}, {Fitzgerald},
  {Fleming}, {Font-Ribera}, {Ford}, {Frinchaboy}, {Garc{\'\i}a P{\'e}rez},
  {Gaudi}, {Ge}, {Ghezzi}, {Gillespie}, {Gilmore}, {Girardi}, {Gott}, {Gould},
  {Grebel}, {Gunn}, {Hamilton}, {Harding}, {Harris}, {Hawley}, {Hearty},
  {Hennawi}, {Gonz{\'a}lez Hern{\'a}ndez}, {Ho}, {Hogg}, {Holtzman},
  {Honscheid}, {Inada}, {Ivans}, {Jiang}, {Jiang}, {Johnson}, {Jordan},
  {Jordan}, {Kauffmann}, {Kazin}, {Kirkby}, {Klaene}, {Knapp}, {Kneib},
  {Kochanek}, {Koesterke}, {Kollmeier}, {Kron}, {Lampeitl}, {Lang}, {Lawler},
  {Le Goff}, {Lee}, {Lee}, {Leisenring}, {Lin}, {Liu}, {Long}, {Loomis},
  {Lucatello}, {Lundgren}, {Lupton}, {Ma}, {Ma}, {MacDonald}, {Mack},
  {Mahadevan}, {Maia}, {Majewski}, {Makler}, {Malanushenko}, {Malanushenko},
  {Mandelbaum}, {Maraston}, {Margala}, {Maseman}, {Masters}, {McBride},
  {McDonald}, {McGreer}, {McMahon}, {Mena Requejo}, {M{\'e}nard},
  {Miralda-Escud{\'e}}, {Morrison}, {Mullally}, {Muna}, {Murayama}, {Myers},
  {Naugle}, {Neto}, {Nguyen}, {Nichol}, {Nidever}, {O'Connell}, {Ogando},
  {Olmstead}, {Oravetz}, {Padmanabhan}, {Paegert}, {Palanque-Delabrouille},
  {Pan}, {Pandey}, {Parejko}, {P{\^a}ris}, {Pellegrini}, {Pepper}, {Percival},
  {Petitjean}, {Pfaffenberger}, {Pforr}, {Phleps}, {Pichon}, {Pieri}, {Prada},
  {Price-Whelan}, {Raddick}, {Ramos}, {Reid}, {Reyle}, {Rich}, {Richards},
  {Rieke}, {Rieke}, {Rix}, {Robin}, {Rocha-Pinto}, {Rockosi}, {Roe},
  {Rollinde}, {Ross}, {Ross}, {Rossetto}, {S{\'a}nchez}, {Santiago}, {Sayres},
  {Schiavon}, {Schlegel}, {Schlesinger}, {Schmidt}, {Schneider}, {Sellgren},
  {Shelden}, {Sheldon}, {Shetrone}, {Shu}, {Silverman}, {Simmerer}, {Simmons},
  {Sivarani}, {Skrutskie}, {Slosar}, {Smee}, {Smith}, {Snedden}, {Stassun},
  {Steele}, {Steinmetz}, {Stockett}, {Stollberg}, {Strauss}, {Szalay},
  {Tanaka}, {Thakar}, {Thomas}, {Tinker}, {Tofflemire}, {Tojeiro}, {Tremonti},
  {Vargas Maga{\~n}a}, {Verde}, {Vogt}, {Wake}, {Wan}, {Wang}, {Weaver},
  {White}, {White}, {Wilson}, {Wisniewski}, {Wood-Vasey}, {Yanny}, {Yasuda},
  {Y{\`e}che}, {York}, {Young}, {Zasowski}, {Zehavi}, \&
  {Zhao}}]{eisenstein2011sdss}
{Eisenstein}, D.~J., {Weinberg}, D.~H., {Agol}, E., {et~al.} 2011, \aj, 142,
  72, \dodoi{10.1088/0004-6256/142/3/72}

\bibitem[{{Favata} {et~al.}(2004){Favata}, {Hughes}, \& {Holz}}]{Favata2004}
{Favata}, M., {Hughes}, S.~A., \& {Holz}, D.~E. 2004, \apjl, 607, L5,
  \dodoi{10.1086/421552}

\bibitem[{{Fitchett}(1983)}]{Fitchett1983}
{Fitchett}, M.~J. 1983, \mnras, 203, 1049, \dodoi{10.1093/mnras/203.4.1049}

\bibitem[{{Gaia Collaboration} {et~al.}(2016){Gaia Collaboration}, {Prusti},
  {de Bruijne}, {Brown}, {Vallenari}, {Babusiaux}, {Bailer-Jones}, {Bastian},
  {Biermann}, {Evans}, {Eyer}, {Jansen}, {Jordi}, {Klioner}, {Lammers},
  {Lindegren}, {Luri}, {Mignard}, {Milligan}, {Panem}, {Poinsignon},
  {Pourbaix}, {Randich}, {Sarri}, {Sartoretti}, {Siddiqui}, {Soubiran},
  {Valette}, {van Leeuwen}, {Walton}, {Aerts}, {Arenou}, {Cropper}, {Drimmel},
  {H{\o}g}, {Katz}, {Lattanzi}, {O'Mullane}, {Grebel}, {Holland}, {Huc},
  {Passot}, {Bramante}, {Cacciari}, {Casta{\~n}eda}, {Chaoul}, {Cheek}, {De
  Angeli}, {Fabricius}, {Guerra}, {Hern{\'a}ndez}, {Jean-Antoine-Piccolo},
  {Masana}, {Messineo}, {Mowlavi}, {Nienartowicz}, {Ord{\'o}{\~n}ez-Blanco},
  {Panuzzo}, {Portell}, {Richards}, {Riello}, {Seabroke}, {Tanga},
  {Th{\'e}venin}, {Torra}, {Els}, {Gracia-Abril}, {Comoretto},
  {Garcia-Reinaldos}, {Lock}, {Mercier}, {Altmann}, {Andrae}, {Astraatmadja},
  {Bellas-Velidis}, {Benson}, {Berthier}, {Blomme}, {Busso}, {Carry},
  {Cellino}, {Clementini}, {Cowell}, {Creevey}, {Cuypers}, {Davidson}, {De
  Ridder}, {de Torres}, {Delchambre}, {Dell'Oro}, {Ducourant}, {Fr{\'e}mat},
  {Garc{\'\i}a-Torres}, {Gosset}, {Halbwachs}, {Hambly}, {Harrison}, {Hauser},
  {Hestroffer}, {Hodgkin}, {Huckle}, {Hutton}, {Jasniewicz}, {Jordan},
  {Kontizas}, {Korn}, {Lanzafame}, {Manteiga}, {Moitinho}, {Muinonen},
  {Osinde}, {Pancino}, {Pauwels}, {Petit}, {Recio-Blanco}, {Robin}, {Sarro},
  {Siopis}, {Smith}, {Smith}, {Sozzetti}, {Thuillot}, {van Reeven}, {Viala},
  {Abbas}, {Abreu Aramburu}, {Accart}, {Aguado}, {Allan}, {Allasia},
  {Altavilla}, {{\'A}lvarez}, {Alves}, {Anderson}, {Andrei}, {Anglada Varela},
  {Antiche}, {Antoja}, {Ant{\'o}n}, {Arcay}, {Atzei}, {Ayache}, {Bach},
  {Baker}, {Balaguer-N{\'u}{\~n}ez}, {Barache}, {Barata}, {Barbier}, {Barblan},
  {Baroni}, {Barrado y Navascu{\'e}s}, {Barros}, {Barstow}, {Becciani},
  {Bellazzini}, {Bellei}, {Bello Garc{\'\i}a}, {Belokurov}, {Bendjoya},
  {Berihuete}, {Bianchi}, {Bienaym{\'e}}, {Billebaud}, {Blagorodnova},
  {Blanco-Cuaresma}, {Boch}, {Bombrun}, {Borrachero}, {Bouquillon}, {Bourda},
  {Bouy}, {Bragaglia}, {Breddels}, {Brouillet}, {Br{\"u}semeister},
  {Bucciarelli}, {Budnik}, {Burgess}, {Burgon}, {Burlacu}, {Busonero}, {Buzzi},
  {Caffau}, {Cambras}, {Campbell}, {Cancelliere}, {Cantat-Gaudin}, {Carlucci},
  {Carrasco}, {Castellani}, {Charlot}, {Charnas}, {Charvet}, {Chassat},
  {Chiavassa}, {Clotet}, {Cocozza}, {Collins}, {Collins}, {Costigan}, {Crifo},
  {Cross}, {Crosta}, {Crowley}, {Dafonte}, {Damerdji}, {Dapergolas}, {David},
  {David}, {De Cat}, {de Felice}, {de Laverny}, {De Luise}, {De March}, {de
  Martino}, {de Souza}, {Debosscher}, {del Pozo}, {Delbo}, {Delgado},
  {Delgado}, {di Marco}, {Di Matteo}, {Diakite}, {Distefano}, {Dolding}, {Dos
  Anjos}, {Drazinos}, {Dur{\'a}n}, {Dzigan}, {Ecale}, {Edvardsson}, {Enke},
  {Erdmann}, {Escolar}, {Espina}, {Evans}, {Eynard Bontemps}, {Fabre},
  {Fabrizio}, {Faigler}, {Falc{\~a}o}, {Farr{\`a}s Casas}, {Faye}, {Federici},
  {Fedorets}, {Fern{\'a}ndez-Hern{\'a}ndez}, {Fernique}, {Fienga}, {Figueras},
  {Filippi}, {Findeisen}, {Fonti}, {Fouesneau}, {Fraile}, {Fraser}, {Fuchs},
  {Furnell}, {Gai}, {Galleti}, {Galluccio}, {Garabato}, {Garc{\'\i}a-Sedano},
  {Gar{\'e}}, {Garofalo}, {Garralda}, {Gavras}, {Gerssen}, {Geyer}, {Gilmore},
  {Girona}, {Giuffrida}, {Gomes}, {Gonz{\'a}lez-Marcos},
  {Gonz{\'a}lez-N{\'u}{\~n}ez}, {Gonz{\'a}lez-Vidal}, {Granvik}, {Guerrier},
  {Guillout}, {Guiraud}, {G{\'u}rpide}, {Guti{\'e}rrez-S{\'a}nchez}, {Guy},
  {Haigron}, {Hatzidimitriou}, {Haywood}, {Heiter}, {Helmi}, {Hobbs},
  {Hofmann}, {Holl}, {Holland}, {Hunt}, {Hypki}, {Icardi}, {Irwin}, {Jevardat
  de Fombelle}, {Jofr{\'e}}, {Jonker}, {Jorissen}, {Julbe}, {Karampelas},
  {Kochoska}, {Kohley}, {Kolenberg}, {Kontizas}, {Koposov}, {Kordopatis},
  {Koubsky}, {Kowalczyk}, {Krone-Martins}, {Kudryashova}, {Kull}, {Bachchan},
  {Lacoste-Seris}, {Lanza}, {Lavigne}, {Le Poncin-Lafitte}, {Lebreton},
  {Lebzelter}, {Leccia}, {Leclerc}, {Lecoeur-Taibi}, {Lemaitre}, {Lenhardt},
  {Leroux}, {Liao}, {Licata}, {Lindstr{\o}m}, {Lister}, {Livanou}, {Lobel},
  {L{\"o}ffler}, {L{\'o}pez}, {Lopez-Lozano}, {Lorenz}, {Loureiro},
  {MacDonald}, {Magalh{\~a}es Fernandes}, {Managau}, {Mann}, {Mantelet},
  {Marchal}, {Marchant}, {Marconi}, {Marie}, {Marinoni}, {Marrese},
  {Marschalk{\'o}}, {Marshall}, {Mart{\'\i}n-Fleitas}, {Martino}, {Mary},
  {Matijevi{\v{c}}}, {Mazeh}, {McMillan}, {Messina}, {Mestre}, {Michalik},
  {Millar}, {Miranda}, {Molina}, {Molinaro}, {Molinaro}, {Moln{\'a}r},
  {Moniez}, {Montegriffo}, {Monteiro}, {Mor}, {Mora}, {Morbidelli}, {Morel},
  {Morgenthaler}, {Morley}, {Morris}, {Mulone}, {Muraveva}, {Musella},
  {Narbonne}, {Nelemans}, {Nicastro}, {Noval}, {Ord{\'e}novic},
  {Ordieres-Mer{\'e}}, {Osborne}, {Pagani}, {Pagano}, {Pailler}, {Palacin},
  {Palaversa}, {Parsons}, {Paulsen}, {Pecoraro}, {Pedrosa}, {Pentik{\"a}inen},
  {Pereira}, {Pichon}, {Piersimoni}, {Pineau}, {Plachy}, {Plum}, {Poujoulet},
  {Pr{\v{s}}a}, {Pulone}, {Ragaini}, {Rago}, {Rambaux}, {Ramos-Lerate},
  {Ranalli}, {Rauw}, {Read}, {Regibo}, {Renk}, {Reyl{\'e}}, {Ribeiro},
  {Rimoldini}, {Ripepi}, {Riva}, {Rixon}, {Roelens}, {Romero-G{\'o}mez},
  {Rowell}, {Royer}, {Rudolph}, {Ruiz-Dern}, {Sadowski}, {Sagrist{\`a}
  Sell{\'e}s}, {Sahlmann}, {Salgado}, {Salguero}, {Sarasso}, {Savietto},
  {Schnorhk}, {Schultheis}, {Sciacca}, {Segol}, {Segovia}, {Segransan},
  {Serpell}, {Shih}, {Smareglia}, {Smart}, {Smith}, {Solano}, {Solitro},
  {Sordo}, {Soria Nieto}, {Souchay}, {Spagna}, {Spoto}, {Stampa}, {Steele},
  {Steidelm{\"u}ller}, {Stephenson}, {Stoev}, {Suess}, {S{\"u}veges}, {Surdej},
  {Szabados}, {Szegedi-Elek}, {Tapiador}, {Taris}, {Tauran}, {Taylor},
  {Teixeira}, {Terrett}, {Tingley}, {Trager}, {Turon}, {Ulla}, {Utrilla},
  {Valentini}, {van Elteren}, {Van Hemelryck}, {van Leeuwen}, {Varadi},
  {Vecchiato}, {Veljanoski}, {Via}, {Vicente}, {Vogt}, {Voss}, {Votruba},
  {Voutsinas}, {Walmsley}, {Weiler}, {Weingrill}, {Werner}, {Wevers},
  {Whitehead}, {Wyrzykowski}, {Yoldas}, {{\v{Z}}erjal}, {Zucker}, {Zurbach},
  {Zwitter}, {Alecu}, {Allen}, {Allende Prieto}, {Amorim},
  {Anglada-Escud{\'e}}, {Arsenijevic}, {Azaz}, {Balm}, {Beck}, {Bernstein},
  {Bigot}, {Bijaoui}, {Blasco}, {Bonfigli}, {Bono}, {Boudreault}, {Bressan},
  {Brown}, {Brunet}, {Bunclark}, {Buonanno}, {Butkevich}, {Carret}, {Carrion},
  {Chemin}, {Ch{\'e}reau}, {Corcione}, {Darmigny}, {de Boer}, {de Teodoro}, {de
  Zeeuw}, {Delle Luche}, {Domingues}, {Dubath}, {Fodor}, {Fr{\'e}zouls},
  {Fries}, {Fustes}, {Fyfe}, {Gallardo}, {Gallegos}, {Gardiol}, {Gebran},
  {Gomboc}, {G{\'o}mez}, {Grux}, {Gueguen}, {Heyrovsky}, {Hoar}, {Iannicola},
  {Isasi Parache}, {Janotto}, {Joliet}, {Jonckheere}, {Keil}, {Kim},
  {Klagyivik}, {Klar}, {Knude}, {Kochukhov}, {Kolka}, {Kos}, {Kutka}, {Lainey},
  {LeBouquin}, {Liu}, {Loreggia}, {Makarov}, {Marseille}, {Martayan},
  {Martinez-Rubi}, {Massart}, {Meynadier}, {Mignot}, {Munari}, {Nguyen},
  {Nordlander}, {Ocvirk}, {O'Flaherty}, {Olias Sanz}, {Ortiz}, {Osorio},
  {Oszkiewicz}, {Ouzounis}, {Palmer}, {Park}, {Pasquato}, {Peltzer}, {Peralta},
  {P{\'e}turaud}, {Pieniluoma}, {Pigozzi}, {Poels}, {Prat}, {Prod'homme},
  {Raison}, {Rebordao}, {Risquez}, {Rocca-Volmerange}, {Rosen}, {Ruiz-Fuertes},
  {Russo}, {Sembay}, {Serraller Vizcaino}, {Short}, {Siebert}, {Silva},
  {Sinachopoulos}, {Slezak}, {Soffel}, {Sosnowska}, {Strai{\v{z}}ys}, {ter
  Linden}, {Terrell}, {Theil}, {Tiede}, {Troisi}, {Tsalmantza}, {Tur},
  {Vaccari}, {Vachier}, {Valles}, {Van Hamme}, {Veltz}, {Virtanen}, {Wallut},
  {Wichmann}, {Wilkinson}, {Ziaeepour}, \& {Zschocke}}]{gaia2016gaia}
{Gaia Collaboration}, {Prusti}, T., {de Bruijne}, J.~H.~J., {et~al.} 2016,
  \aap, 595, A1, \dodoi{10.1051/0004-6361/201629272}

\bibitem[{{Gaia Collaboration} {et~al.}(2018){Gaia Collaboration}, {Brown},
  {Vallenari}, {Prusti}, {de Bruijne}, {Babusiaux}, {Bailer-Jones}, {Biermann},
  {Evans}, {Eyer}, {Jansen}, {Jordi}, {Klioner}, {Lammers}, {Lindegren},
  {Luri}, {Mignard}, {Panem}, {Pourbaix}, {Randich}, {Sartoretti}, {Siddiqui},
  {Soubiran}, {van Leeuwen}, {Walton}, {Arenou}, {Bastian}, {Cropper},
  {Drimmel}, {Katz}, {Lattanzi}, {Bakker}, {Cacciari}, {Casta{\~n}eda},
  {Chaoul}, {Cheek}, {De Angeli}, {Fabricius}, {Guerra}, {Holl}, {Masana},
  {Messineo}, {Mowlavi}, {Nienartowicz}, {Panuzzo}, {Portell}, {Riello},
  {Seabroke}, {Tanga}, {Th{\'e}venin}, {Gracia-Abril}, {Comoretto},
  {Garcia-Reinaldos}, {Teyssier}, {Altmann}, {Andrae}, {Audard},
  {Bellas-Velidis}, {Benson}, {Berthier}, {Blomme}, {Burgess}, {Busso},
  {Carry}, {Cellino}, {Clementini}, {Clotet}, {Creevey}, {Davidson}, {De
  Ridder}, {Delchambre}, {Dell'Oro}, {Ducourant},
  {Fern{\'a}ndez-Hern{\'a}ndez}, {Fouesneau}, {Fr{\'e}mat}, {Galluccio},
  {Garc{\'\i}a-Torres}, {Gonz{\'a}lez-N{\'u}{\~n}ez}, {Gonz{\'a}lez-Vidal},
  {Gosset}, {Guy}, {Halbwachs}, {Hambly}, {Harrison}, {Hern{\'a}ndez},
  {Hestroffer}, {Hodgkin}, {Hutton}, {Jasniewicz}, {Jean-Antoine-Piccolo},
  {Jordan}, {Korn}, {Krone-Martins}, {Lanzafame}, {Lebzelter}, {L{\"o}ffler},
  {Manteiga}, {Marrese}, {Mart{\'\i}n-Fleitas}, {Moitinho}, {Mora}, {Muinonen},
  {Osinde}, {Pancino}, {Pauwels}, {Petit}, {Recio-Blanco}, {Richards},
  {Rimoldini}, {Robin}, {Sarro}, {Siopis}, {Smith}, {Sozzetti}, {S{\"u}veges},
  {Torra}, {van Reeven}, {Abbas}, {Abreu Aramburu}, {Accart}, {Aerts},
  {Altavilla}, {{\'A}lvarez}, {Alvarez}, {Alves}, {Anderson}, {Andrei},
  {Anglada Varela}, {Antiche}, {Antoja}, {Arcay}, {Astraatmadja}, {Bach},
  {Baker}, {Balaguer-N{\'u}{\~n}ez}, {Balm}, {Barache}, {Barata}, {Barbato},
  {Barblan}, {Barklem}, {Barrado}, {Barros}, {Barstow}, {Bartholom{\'e}
  Mu{\~n}oz}, {Bassilana}, {Becciani}, {Bellazzini}, {Berihuete}, {Bertone},
  {Bianchi}, {Bienaym{\'e}}, {Blanco-Cuaresma}, {Boch}, {Boeche}, {Bombrun},
  {Borrachero}, {Bossini}, {Bouquillon}, {Bourda}, {Bragaglia}, {Bramante},
  {Breddels}, {Bressan}, {Brouillet}, {Br{\"u}semeister}, {Brugaletta},
  {Bucciarelli}, {Burlacu}, {Busonero}, {Butkevich}, {Buzzi}, {Caffau},
  {Cancelliere}, {Cannizzaro}, {Cantat-Gaudin}, {Carballo}, {Carlucci},
  {Carrasco}, {Casamiquela}, {Castellani}, {Castro-Ginard}, {Charlot},
  {Chemin}, {Chiavassa}, {Cocozza}, {Costigan}, {Cowell}, {Crifo}, {Crosta},
  {Crowley}, {Cuypers}, {Dafonte}, {Damerdji}, {Dapergolas}, {David}, {David},
  {de Laverny}, {De Luise}, {De March}, {de Martino}, {de Souza}, {de Torres},
  {Debosscher}, {del Pozo}, {Delbo}, {Delgado}, {Delgado}, {Di Matteo},
  {Diakite}, {Diener}, {Distefano}, {Dolding}, {Drazinos}, {Dur{\'a}n},
  {Edvardsson}, {Enke}, {Eriksson}, {Esquej}, {Eynard Bontemps}, {Fabre},
  {Fabrizio}, {Faigler}, {Falc{\~a}o}, {Farr{\`a}s Casas}, {Federici},
  {Fedorets}, {Fernique}, {Figueras}, {Filippi}, {Findeisen}, {Fonti},
  {Fraile}, {Fraser}, {Fr{\'e}zouls}, {Gai}, {Galleti}, {Garabato},
  {Garc{\'\i}a-Sedano}, {Garofalo}, {Garralda}, {Gavel}, {Gavras}, {Gerssen},
  {Geyer}, {Giacobbe}, {Gilmore}, {Girona}, {Giuffrida}, {Glass}, {Gomes},
  {Granvik}, {Gueguen}, {Guerrier}, {Guiraud}, {Guti{\'e}rrez-S{\'a}nchez},
  {Haigron}, {Hatzidimitriou}, {Hauser}, {Haywood}, {Heiter}, {Helmi}, {Heu},
  {Hilger}, {Hobbs}, {Hofmann}, {Holland}, {Huckle}, {Hypki}, {Icardi},
  {Jan{\ss}en}, {Jevardat de Fombelle}, {Jonker}, {Juh{\'a}sz}, {Julbe},
  {Karampelas}, {Kewley}, {Klar}, {Kochoska}, {Kohley}, {Kolenberg},
  {Kontizas}, {Kontizas}, {Koposov}, {Kordopatis}, {Kostrzewa-Rutkowska},
  {Koubsky}, {Lambert}, {Lanza}, {Lasne}, {Lavigne}, {Le Fustec}, {Le
  Poncin-Lafitte}, {Lebreton}, {Leccia}, {Leclerc}, {Lecoeur-Taibi},
  {Lenhardt}, {Leroux}, {Liao}, {Licata}, {Lindstr{\o}m}, {Lister}, {Livanou},
  {Lobel}, {L{\'o}pez}, {Managau}, {Mann}, {Mantelet}, {Marchal}, {Marchant},
  {Marconi}, {Marinoni}, {Marschalk{\'o}}, {Marshall}, {Martino}, {Marton},
  {Mary}, {Massari}, {Matijevi{\v{c}}}, {Mazeh}, {McMillan}, {Messina},
  {Michalik}, {Millar}, {Molina}, {Molinaro}, {Moln{\'a}r}, {Montegriffo},
  {Mor}, {Morbidelli}, {Morel}, {Morris}, {Mulone}, {Muraveva}, {Musella},
  {Nelemans}, {Nicastro}, {Noval}, {O'Mullane}, {Ord{\'e}novic},
  {Ord{\'o}{\~n}ez-Blanco}, {Osborne}, {Pagani}, {Pagano}, {Pailler},
  {Palacin}, {Palaversa}, {Panahi}, {Pawlak}, {Piersimoni}, {Pineau}, {Plachy},
  {Plum}, {Poggio}, {Poujoulet}, {Pr{\v{s}}a}, {Pulone}, {Racero}, {Ragaini},
  {Rambaux}, {Ramos-Lerate}, {Regibo}, {Reyl{\'e}}, {Riclet}, {Ripepi}, {Riva},
  {Rivard}, {Rixon}, {Roegiers}, {Roelens}, {Romero-G{\'o}mez}, {Rowell},
  {Royer}, {Ruiz-Dern}, {Sadowski}, {Sagrist{\`a} Sell{\'e}s}, {Sahlmann},
  {Salgado}, {Salguero}, {Sanna}, {Santana-Ros}, {Sarasso}, {Savietto},
  {Schultheis}, {Sciacca}, {Segol}, {Segovia}, {S{\'e}gransan}, {Shih},
  {Siltala}, {Silva}, {Smart}, {Smith}, {Solano}, {Solitro}, {Sordo}, {Soria
  Nieto}, {Souchay}, {Spagna}, {Spoto}, {Stampa}, {Steele},
  {Steidelm{\"u}ller}, {Stephenson}, {Stoev}, {Suess}, {Surdej}, {Szabados},
  {Szegedi-Elek}, {Tapiador}, {Taris}, {Tauran}, {Taylor}, {Teixeira},
  {Terrett}, {Teyssandier}, {Thuillot}, {Titarenko}, {Torra Clotet}, {Turon},
  {Ulla}, {Utrilla}, {Uzzi}, {Vaillant}, {Valentini}, {Valette}, {van Elteren},
  {Van Hemelryck}, {van Leeuwen}, {Vaschetto}, {Vecchiato}, {Veljanoski},
  {Viala}, {Vicente}, {Vogt}, {von Essen}, {Voss}, {Votruba}, {Voutsinas},
  {Walmsley}, {Weiler}, {Wertz}, {Wevers}, {Wyrzykowski}, {Yoldas},
  {{\v{Z}}erjal}, {Ziaeepour}, {Zorec}, {Zschocke}, {Zucker}, {Zurbach}, \&
  {Zwitter}}]{Gaiadr2}
{Gaia Collaboration}, {Brown}, A.~G.~A., {Vallenari}, A., {et~al.} 2018, \aap,
  616, A1, \dodoi{10.1051/0004-6361/201833051}

\bibitem[{{Gaia Collaboration} {et~al.}(2021){Gaia Collaboration}, {Brown},
  {Vallenari}, {Prusti}, {de Bruijne}, {Babusiaux}, {Biermann}, {Creevey},
  {Evans}, {Eyer}, {Hutton}, {Jansen}, {Jordi}, {Klioner}, {Lammers},
  {Lindegren}, {Luri}, {Mignard}, {Panem}, {Pourbaix}, {Randich}, {Sartoretti},
  {Soubiran}, {Walton}, {Arenou}, {Bailer-Jones}, {Bastian}, {Cropper},
  {Drimmel}, {Katz}, {Lattanzi}, {van Leeuwen}, {Bakker}, {Cacciari},
  {Casta{\~n}eda}, {De Angeli}, {Ducourant}, {Fabricius}, {Fouesneau},
  {Fr{\'e}mat}, {Guerra}, {Guerrier}, {Guiraud}, {Jean-Antoine Piccolo},
  {Masana}, {Messineo}, {Mowlavi}, {Nicolas}, {Nienartowicz}, {Pailler},
  {Panuzzo}, {Riclet}, {Roux}, {Seabroke}, {Sordo}, {Tanga}, {Th{\'e}venin},
  {Gracia-Abril}, {Portell}, {Teyssier}, {Altmann}, {Andrae}, {Bellas-Velidis},
  {Benson}, {Berthier}, {Blomme}, {Brugaletta}, {Burgess}, {Busso}, {Carry},
  {Cellino}, {Cheek}, {Clementini}, {Damerdji}, {Davidson}, {Delchambre},
  {Dell'Oro}, {Fern{\'a}ndez-Hern{\'a}ndez}, {Galluccio}, {Garc{\'\i}a-Lario},
  {Garcia-Reinaldos}, {Gonz{\'a}lez-N{\'u}{\~n}ez}, {Gosset}, {Haigron},
  {Halbwachs}, {Hambly}, {Harrison}, {Hatzidimitriou}, {Heiter},
  {Hern{\'a}ndez}, {Hestroffer}, {Hodgkin}, {Holl}, {Jan{\ss}en}, {Jevardat de
  Fombelle}, {Jordan}, {Krone-Martins}, {Lanzafame}, {L{\"o}ffler}, {Lorca},
  {Manteiga}, {Marchal}, {Marrese}, {Moitinho}, {Mora}, {Muinonen}, {Osborne},
  {Pancino}, {Pauwels}, {Petit}, {Recio-Blanco}, {Richards}, {Riello},
  {Rimoldini}, {Robin}, {Roegiers}, {Rybizki}, {Sarro}, {Siopis}, {Smith},
  {Sozzetti}, {Ulla}, {Utrilla}, {van Leeuwen}, {van Reeven}, {Abbas}, {Abreu
  Aramburu}, {Accart}, {Aerts}, {Aguado}, {Ajaj}, {Altavilla}, {{\'A}lvarez},
  {{\'A}lvarez Cid-Fuentes}, {Alves}, {Anderson}, {Anglada Varela}, {Antoja},
  {Audard}, {Baines}, {Baker}, {Balaguer-N{\'u}{\~n}ez}, {Balbinot}, {Balog},
  {Barache}, {Barbato}, {Barros}, {Barstow}, {Bartolom{\'e}}, {Bassilana},
  {Bauchet}, {Baudesson-Stella}, {Becciani}, {Bellazzini}, {Bernet}, {Bertone},
  {Bianchi}, {Blanco-Cuaresma}, {Boch}, {Bombrun}, {Bossini}, {Bouquillon},
  {Bragaglia}, {Bramante}, {Breedt}, {Bressan}, {Brouillet}, {Bucciarelli},
  {Burlacu}, {Busonero}, {Butkevich}, {Buzzi}, {Caffau}, {Cancelliere},
  {C{\'a}novas}, {Cantat-Gaudin}, {Carballo}, {Carlucci}, {Carnerero},
  {Carrasco}, {Casamiquela}, {Castellani}, {Castro-Ginard}, {Castro Sampol},
  {Chaoul}, {Charlot}, {Chemin}, {Chiavassa}, {Cioni}, {Comoretto}, {Cooper},
  {Cornez}, {Cowell}, {Crifo}, {Crosta}, {Crowley}, {Dafonte}, {Dapergolas},
  {David}, {David}, {de Laverny}, {De Luise}, {De March}, {De Ridder}, {de
  Souza}, {de Teodoro}, {de Torres}, {del Peloso}, {del Pozo}, {Delbo},
  {Delgado}, {Delgado}, {Delisle}, {Di Matteo}, {Diakite}, {Diener},
  {Distefano}, {Dolding}, {Eappachen}, {Edvardsson}, {Enke}, {Esquej}, {Fabre},
  {Fabrizio}, {Faigler}, {Fedorets}, {Fernique}, {Fienga}, {Figueras},
  {Fouron}, {Fragkoudi}, {Fraile}, {Franke}, {Gai}, {Garabato},
  {Garcia-Gutierrez}, {Garc{\'\i}a-Torres}, {Garofalo}, {Gavras}, {Gerlach},
  {Geyer}, {Giacobbe}, {Gilmore}, {Girona}, {Giuffrida}, {Gomel}, {Gomez},
  {Gonzalez-Santamaria}, {Gonz{\'a}lez-Vidal}, {Granvik},
  {Guti{\'e}rrez-S{\'a}nchez}, {Guy}, {Hauser}, {Haywood}, {Helmi}, {Hidalgo},
  {Hilger}, {H{\l}adczuk}, {Hobbs}, {Holland}, {Huckle}, {Jasniewicz},
  {Jonker}, {Juaristi Campillo}, {Julbe}, {Karbevska}, {Kervella}, {Khanna},
  {Kochoska}, {Kontizas}, {Kordopatis}, {Korn}, {Kostrzewa-Rutkowska},
  {Kruszy{\'n}ska}, {Lambert}, {Lanza}, {Lasne}, {Le Campion}, {Le Fustec},
  {Lebreton}, {Lebzelter}, {Leccia}, {Leclerc}, {Lecoeur-Taibi}, {Liao},
  {Licata}, {Lindstr{\o}m}, {Lister}, {Livanou}, {Lobel}, {Madrero Pardo},
  {Managau}, {Mann}, {Marchant}, {Marconi}, {Marcos Santos}, {Marinoni},
  {Marocco}, {Marshall}, {Martin Polo}, {Mart{\'\i}n-Fleitas}, {Masip},
  {Massari}, {Mastrobuono-Battisti}, {Mazeh}, {McMillan}, {Messina},
  {Michalik}, {Millar}, {Mints}, {Molina}, {Molinaro}, {Moln{\'a}r},
  {Montegriffo}, {Mor}, {Morbidelli}, {Morel}, {Morris}, {Mulone}, {Munoz},
  {Muraveva}, {Murphy}, {Musella}, {Noval}, {Ord{\'e}novic}, {Orr{\`u}},
  {Osinde}, {Pagani}, {Pagano}, {Palaversa}, {Palicio}, {Panahi}, {Pawlak},
  {Pe{\~n}alosa Esteller}, {Penttil{\"a}}, {Piersimoni}, {Pineau}, {Plachy},
  {Plum}, {Poggio}, {Poretti}, {Poujoulet}, {Pr{\v{s}}a}, {Pulone}, {Racero},
  {Ragaini}, {Rainer}, {Raiteri}, {Rambaux}, {Ramos}, {Ramos-Lerate}, {Re
  Fiorentin}, {Regibo}, {Reyl{\'e}}, {Ripepi}, {Riva}, {Rixon}, {Robichon},
  {Robin}, {Roelens}, {Rohrbasser}, {Romero-G{\'o}mez}, {Rowell}, {Royer},
  {Rybicki}, {Sadowski}, {Sagrist{\`a} Sell{\'e}s}, {Sahlmann}, {Salgado},
  {Salguero}, {Samaras}, {Sanchez Gimenez}, {Sanna}, {Santove{\~n}a},
  {Sarasso}, {Schultheis}, {Sciacca}, {Segol}, {Segovia}, {S{\'e}gransan},
  {Semeux}, {Shahaf}, {Siddiqui}, {Siebert}, {Siltala}, {Slezak}, {Smart},
  {Solano}, {Solitro}, {Souami}, {Souchay}, {Spagna}, {Spoto}, {Steele},
  {Steidelm{\"u}ller}, {Stephenson}, {S{\"u}veges}, {Szabados}, {Szegedi-Elek},
  {Taris}, {Tauran}, {Taylor}, {Teixeira}, {Thuillot}, {Tonello}, {Torra},
  {Torra}, {Turon}, {Unger}, {Vaillant}, {van Dillen}, {Vanel}, {Vecchiato},
  {Viala}, {Vicente}, {Voutsinas}, {Weiler}, {Wevers}, {Wyrzykowski}, {Yoldas},
  {Yvard}, {Zhao}, {Zorec}, {Zucker}, {Zurbach}, \& {Zwitter}}]{Gaiaedr3}
---. 2021, \aap, 649, A1, \dodoi{10.1051/0004-6361/202039657}

\bibitem[{{Greene} {et~al.}(2021){Greene}, {Lancaster}, {Ting}, {Koposov},
  {Danieli}, {Huang}, {Jiang}, {Greco}, \& {Strader}}]{greene2021search}
{Greene}, J.~E., {Lancaster}, L., {Ting}, Y.-S., {et~al.} 2021, \apj, 917, 17,
  \dodoi{10.3847/1538-4357/ac0896}

\bibitem[{Hobbs {et~al.}(2022)Hobbs, Lindegren, Bastian, Klioner, Butkevich,
  Stephenson, Hernandez, Lammers, Bombrun, Mignard, {et~al.}}]{hobbs2022gaia}
Hobbs, D., Lindegren, L., Bastian, U., {et~al.} 2022, Gaia DR3 documentation, 4

\bibitem[{{Hughes} {et~al.}(2021){Hughes}, {Sand}, {Seth}, {Strader}, {Voggel},
  {Dumont}, {Crnojevi{\'c}}, {Caldwell}, {Forbes}, {Simon}, {Guhathakurta}, \&
  {Toloba}}]{hughes2021ngc}
{Hughes}, A.~K., {Sand}, D.~J., {Seth}, A., {et~al.} 2021, \apj, 914, 16,
  \dodoi{10.3847/1538-4357/abf63c}

\bibitem[{{Jordi} {et~al.}(2010){Jordi}, {Gebran}, {Carrasco}, {de Bruijne},
  {Voss}, {Fabricius}, {Knude}, {Vallenari}, {Kohley}, \& {Mora}}]{Jordi2010}
{Jordi}, C., {Gebran}, M., {Carrasco}, J.~M., {et~al.} 2010, \aap, 523, A48,
  \dodoi{10.1051/0004-6361/201015441}

\bibitem[{{Komossa} \& {Merritt}(2008)}]{Komossa2008}
{Komossa}, S., \& {Merritt}, D. 2008, \apjl, 683, L21, \dodoi{10.1086/591420}

\bibitem[{{Laureijs} {et~al.}(2011){Laureijs}, {Amiaux}, {Arduini},
  {Augu{\`e}res}, {Brinchmann}, {Cole}, {Cropper}, {Dabin}, {Duvet}, {Ealet},
  {Garilli}, {Gondoin}, {Guzzo}, {Hoar}, {Hoekstra}, {Holmes}, {Kitching},
  {Maciaszek}, {Mellier}, {Pasian}, {Percival}, {Rhodes}, {Saavedra Criado},
  {Sauvage}, {Scaramella}, {Valenziano}, {Warren}, {Bender}, {Castander},
  {Cimatti}, {Le F{\`e}vre}, {Kurki-Suonio}, {Levi}, {Lilje}, {Meylan},
  {Nichol}, {Pedersen}, {Popa}, {Rebolo Lopez}, {Rix}, {Rottgering},
  {Zeilinger}, {Grupp}, {Hudelot}, {Massey}, {Meneghetti}, {Miller}, {Paltani},
  {Paulin-Henriksson}, {Pires}, {Saxton}, {Schrabback}, {Seidel}, {Walsh},
  {Aghanim}, {Amendola}, {Bartlett}, {Baccigalupi}, {Beaulieu}, {Benabed},
  {Cuby}, {Elbaz}, {Fosalba}, {Gavazzi}, {Helmi}, {Hook}, {Irwin}, {Kneib},
  {Kunz}, {Mannucci}, {Moscardini}, {Tao}, {Teyssier}, {Weller}, {Zamorani},
  {Zapatero Osorio}, {Boulade}, {Foumond}, {Di Giorgio}, {Guttridge}, {James},
  {Kemp}, {Martignac}, {Spencer}, {Walton}, {Bl{\"u}mchen}, {Bonoli},
  {Bortoletto}, {Cerna}, {Corcione}, {Fabron}, {Jahnke}, {Ligori}, {Madrid},
  {Martin}, {Morgante}, {Pamplona}, {Prieto}, {Riva}, {Toledo}, {Trifoglio},
  {Zerbi}, {Abdalla}, {Douspis}, {Grenet}, {Borgani}, {Bouwens}, {Courbin},
  {Delouis}, {Dubath}, {Fontana}, {Frailis}, {Grazian}, {Koppenh{\"o}fer},
  {Mansutti}, {Melchior}, {Mignoli}, {Mohr}, {Neissner}, {Noddle}, {Poncet},
  {Scodeggio}, {Serrano}, {Shane}, {Starck}, {Surace}, {Taylor},
  {Verdoes-Kleijn}, {Vuerli}, {Williams}, {Zacchei}, {Altieri}, {Escudero
  Sanz}, {Kohley}, {Oosterbroek}, {Astier}, {Bacon}, {Bardelli}, {Baugh},
  {Bellagamba}, {Benoist}, {Bianchi}, {Biviano}, {Branchini}, {Carbone},
  {Cardone}, {Clements}, {Colombi}, {Conselice}, {Cresci}, {Deacon}, {Dunlop},
  {Fedeli}, {Fontanot}, {Franzetti}, {Giocoli}, {Garcia-Bellido}, {Gow},
  {Heavens}, {Hewett}, {Heymans}, {Holland}, {Huang}, {Ilbert}, {Joachimi},
  {Jennins}, {Kerins}, {Kiessling}, {Kirk}, {Kotak}, {Krause}, {Lahav}, {van
  Leeuwen}, {Lesgourgues}, {Lombardi}, {Magliocchetti}, {Maguire}, {Majerotto},
  {Maoli}, {Marulli}, {Maurogordato}, {McCracken}, {McLure}, {Melchiorri},
  {Merson}, {Moresco}, {Nonino}, {Norberg}, {Peacock}, {Pello}, {Penny},
  {Pettorino}, {Di Porto}, {Pozzetti}, {Quercellini}, {Radovich}, {Rassat},
  {Roche}, {Ronayette}, {Rossetti}, {Sartoris}, {Schneider}, {Semboloni},
  {Serjeant}, {Simpson}, {Skordis}, {Smadja}, {Smartt}, {Spano}, {Spiro},
  {Sullivan}, {Tilquin}, {Trotta}, {Verde}, {Wang}, {Williger}, {Zhao},
  {Zoubian}, \& {Zucca}}]{laureijs2011euclid}
{Laureijs}, R., {Amiaux}, J., {Arduini}, S., {et~al.} 2011, arXiv e-prints,
  arXiv:1110.3193, \dodoi{10.48550/arXiv.1110.3193}

\bibitem[{{Lena} {et~al.}(2020){Lena}, {Jonker}, {Rauer}, {Hernandez}, \&
  {Kostrzewa-Rutkowska}}]{lena2020hypercompact}
{Lena}, D., {Jonker}, P.~G., {Rauer}, J.~P., {Hernandez}, S., \&
  {Kostrzewa-Rutkowska}, Z. 2020, \mnras, 495, 1771,
  \dodoi{10.1093/mnras/staa1174}

\bibitem[{{Libeskind} {et~al.}(2006){Libeskind}, {Cole}, {Frenk}, \&
  {Helly}}]{libeskind2006effect}
{Libeskind}, N.~I., {Cole}, S., {Frenk}, C.~S., \& {Helly}, J.~C. 2006, \mnras,
  368, 1381, \dodoi{10.1111/j.1365-2966.2006.10209.x}

\bibitem[{{Lindegren} {et~al.}(2012){Lindegren}, {Lammers}, {Hobbs},
  {O'Mullane}, {Bastian}, \& {Hern{\'a}ndez}}]{astrometric_core_solution}
{Lindegren}, L., {Lammers}, U., {Hobbs}, D., {et~al.} 2012, \aap, 538, A78,
  \dodoi{10.1051/0004-6361/201117905}

\bibitem[{{Lindegren} {et~al.}(2008){Lindegren}, {Babusiaux}, {Bailer-Jones},
  {Bastian}, {Brown}, {Cropper}, {H{\o}g}, {Jordi}, {Katz}, {van Leeuwen},
  {Luri}, {Mignard}, {de Bruijne}, \& {Prusti}}]{lindegren2007gaia}
{Lindegren}, L., {Babusiaux}, C., {Bailer-Jones}, C., {et~al.} 2008, in A Giant
  Step: from Milli- to Micro-arcsecond Astrometry, ed. W.~J. {Jin},
  I.~{Platais}, \& M.~A.~C. {Perryman}, Vol. 248, 217--223,
  \dodoi{10.1017/S1743921308019133}

\bibitem[{{Lindegren} {et~al.}(2018){Lindegren}, {Hern{\'a}ndez}, {Bombrun},
  {Klioner}, {Bastian}, {Ramos-Lerate}, {de Torres}, {Steidelm{\"u}ller},
  {Stephenson}, {Hobbs}, {Lammers}, {Biermann}, {Geyer}, {Hilger}, {Michalik},
  {Stampa}, {McMillan}, {Casta{\~n}eda}, {Clotet}, {Comoretto}, {Davidson},
  {Fabricius}, {Gracia}, {Hambly}, {Hutton}, {Mora}, {Portell}, {van Leeuwen},
  {Abbas}, {Abreu}, {Altmann}, {Andrei}, {Anglada}, {Balaguer-N{\'u}{\~n}ez},
  {Barache}, {Becciani}, {Bertone}, {Bianchi}, {Bouquillon}, {Bourda},
  {Br{\"u}semeister}, {Bucciarelli}, {Busonero}, {Buzzi}, {Cancelliere},
  {Carlucci}, {Charlot}, {Cheek}, {Crosta}, {Crowley}, {de Bruijne}, {de
  Felice}, {Drimmel}, {Esquej}, {Fienga}, {Fraile}, {Gai}, {Garralda},
  {Gonz{\'a}lez-Vidal}, {Guerra}, {Hauser}, {Hofmann}, {Holl}, {Jordan},
  {Lattanzi}, {Lenhardt}, {Liao}, {Licata}, {Lister}, {L{\"o}ffler},
  {Marchant}, {Martin-Fleitas}, {Messineo}, {Mignard}, {Morbidelli}, {Poggio},
  {Riva}, {Rowell}, {Salguero}, {Sarasso}, {Sciacca}, {Siddiqui}, {Smart},
  {Spagna}, {Steele}, {Taris}, {Torra}, {van Elteren}, {van Reeven}, \&
  {Vecchiato}}]{lindegren2018gaia}
{Lindegren}, L., {Hern{\'a}ndez}, J., {Bombrun}, A., {et~al.} 2018, \aap, 616,
  A2, \dodoi{10.1051/0004-6361/201832727}

\bibitem[{{Lousto} \& {Zlochower}(2011)}]{Lousto2011}
{Lousto}, C.~O., \& {Zlochower}, Y. 2011, \prl, 107, 231102,
  \dodoi{10.1103/PhysRevLett.107.231102}

\bibitem[{{Lousto} {et~al.}(2012){Lousto}, {Zlochower}, {Dotti}, \&
  {Volonteri}}]{Lousto2012}
{Lousto}, C.~O., {Zlochower}, Y., {Dotti}, M., \& {Volonteri}, M. 2012, \prd,
  85, 084015, \dodoi{10.1103/PhysRevD.85.084015}

\bibitem[{{Luo} {et~al.}(2015){Luo}, {Zhao}, {Zhao}, {Deng}, {Liu}, {Jing},
  {Wang}, {Zhang}, {Shi}, {Cui}, {Chu}, {Li}, {Bai}, {Wu}, {Cai}, {Cao}, {Cao},
  {Carlin}, {Chen}, {Chen}, {Chen}, {Chen}, {Chen}, {Chen}, {Chen},
  {Christlieb}, {Chu}, {Cui}, {Dong}, {Du}, {Fan}, {Feng}, {Fu}, {Gao}, {Gong},
  {Gu}, {Guo}, {Han}, {He}, {Hou}, {Hou}, {Hou}, {Hu}, {Hu}, {Hu}, {Huo},
  {Jia}, {Jiang}, {Jiang}, {Jiang}, {Jin}, {Kong}, {Kong}, {Lei}, {Li}, {Li},
  {Li}, {Li}, {Li}, {Li}, {Li}, {Li}, {Li}, {Li}, {Li}, {Li}, {Liang}, {Lin},
  {Liu}, {Liu}, {Liu}, {Liu}, {Lu}, {Luo}, {Mao}, {Newberg}, {Ni}, {Qi}, {Qi},
  {Shen}, {Shi}, {Song}, {Song}, {Su}, {Su}, {Tang}, {Tao}, {Tian}, {Wang},
  {Wang}, {Wang}, {Wang}, {Wang}, {Wang}, {Wang}, {Wang}, {Wang}, {Wang},
  {Wang}, {Wang}, {Wang}, {Wang}, {Wang}, {Wang}, {Wang}, {Wang}, {Wang},
  {Wang}, {Wei}, {Wei}, {Wu}, {Wu}, {Wu}, {Wu}, {Xing}, {Xu}, {Xu}, {Xu},
  {Yan}, {Yang}, {Yang}, {Yang}, {Yang}, {Yao}, {Yu}, {Yuan}, {Yuan}, {Yuan},
  {Yuan}, {Zhai}, {Zhang}, {Zhang}, {Zhang}, {Zhang}, {Zhang}, {Zhang},
  {Zhang}, {Zhang}, {Zhao}, {Zhou}, {Zhou}, {Zhu}, {Zhu}, {Zou}, \&
  {Zuo}}]{luo2015first}
{Luo}, A.~L., {Zhao}, Y.-H., {Zhao}, G., {et~al.} 2015, Research in Astronomy
  and Astrophysics, 15, 1095, \dodoi{10.1088/1674-4527/15/8/002}

\bibitem[{{Mannucci} {et~al.}(2022){Mannucci}, {Pancino}, {Belfiore}, {Cicone},
  {Ciurlo}, {Cresci}, {Lusso}, {Marasco}, {Marconi}, {Nardini}, {Pinna},
  {Severgnini}, {Saracco}, {Tozzi}, \& {Yeh}}]{mannucci2022unveiling}
{Mannucci}, F., {Pancino}, E., {Belfiore}, F., {et~al.} 2022, Nature Astronomy,
  6, 1185, \dodoi{10.1038/s41550-022-01761-5}

\bibitem[{{Merritt} {et~al.}(2004){Merritt}, {Milosavljevi{\'c}}, {Favata},
  {Hughes}, \& {Holz}}]{Merritt2004}
{Merritt}, D., {Milosavljevi{\'c}}, M., {Favata}, M., {Hughes}, S.~A., \&
  {Holz}, D.~E. 2004, \apjl, 607, L9, \dodoi{10.1086/421551}

\bibitem[{{Merritt} {et~al.}(2009){Merritt}, {Schnittman}, \&
  {Komossa}}]{Merritt2009}
{Merritt}, D., {Schnittman}, J.~D., \& {Komossa}, S. 2009, \apj, 699, 1690,
  \dodoi{10.1088/0004-637X/699/2/1690}

\bibitem[{{Obasi} {et~al.}(2023){Obasi}, {G{\'o}mez}, {Minniti},
  {Alonso-Garc{\'\i}a}, {Hempel}, {Pullen}, {Gregg}, {Baravalle}, {Alonso}, \&
  {Okere}}]{obasi2023globular}
{Obasi}, C., {G{\'o}mez}, M., {Minniti}, D., {et~al.} 2023, \aap, 670, A18,
  \dodoi{10.1051/0004-6361/202243154}

\bibitem[{{O'Leary} \& {Loeb}(2009)}]{o2009star}
{O'Leary}, R.~M., \& {Loeb}, A. 2009, \mnras, 395, 781,
  \dodoi{10.1111/j.1365-2966.2009.14611.x}

\bibitem[{{O'Leary} \& {Loeb}(2012)}]{o2012recoiled}
---. 2012, \mnras, 421, 2737, \dodoi{10.1111/j.1365-2966.2011.20078.x}

\bibitem[{{Oudmaijer} {et~al.}(2022){Oudmaijer}, {Jones}, \&
  {Vioque}}]{oudmaijer2022census}
{Oudmaijer}, R.~D., {Jones}, E. R.~M., \& {Vioque}, M. 2022, \mnras, 516, L61,
  \dodoi{10.1093/mnrasl/slac088}

\bibitem[{{Penoyre} {et~al.}(2022){Penoyre}, {Belokurov}, \&
  {Evans}}]{Penoyre2022}
{Penoyre}, Z., {Belokurov}, V., \& {Evans}, N.~W. 2022, \mnras, 513, 5270,
  \dodoi{10.1093/mnras/stac1147}

\bibitem[{{Peres}(1962)}]{Peres1962}
{Peres}, A. 1962, Physical Review, 128, 2471, \dodoi{10.1103/PhysRev.128.2471}

\bibitem[{Pourbaix {et~al.}(2022)Pourbaix, Arenou, Gavras, Gosset, Halbwachs,
  Siopis, Sozzetti, Bauchet, Damerdji, Delchambre, {et~al.}}]{pourbaix2022gaia}
Pourbaix, D., Arenou, F., Gavras, P., {et~al.} 2022, Gaia DR3 documentation, 7

\bibitem[{{Redmount} \& {Rees}(1989)}]{Redmount1989}
{Redmount}, I.~H., \& {Rees}, M.~J. 1989, Comments on Astrophysics, 14, 165

\bibitem[{{Riello} {et~al.}(2021){Riello}, {De Angeli}, {Evans}, {Montegriffo},
  {Carrasco}, {Busso}, {Palaversa}, {Burgess}, {Diener}, {Davidson}, {Rowell},
  {Fabricius}, {Jordi}, {Bellazzini}, {Pancino}, {Harrison}, {Cacciari}, {van
  Leeuwen}, {Hambly}, {Hodgkin}, {Osborne}, {Altavilla}, {Barstow}, {Brown},
  {Castellani}, {Cowell}, {De Luise}, {Gilmore}, {Giuffrida}, {Hidalgo},
  {Holland}, {Marinoni}, {Pagani}, {Piersimoni}, {Pulone}, {Ragaini}, {Rainer},
  {Richards}, {Sanna}, {Walton}, {Weiler}, \& {Yoldas}}]{riello2021gaia}
{Riello}, M., {De Angeli}, F., {Evans}, D.~W., {et~al.} 2021, \aap, 649, A3,
  \dodoi{10.1051/0004-6361/202039587}

\bibitem[{{Rowell} {et~al.}(2021){Rowell}, {Davidson}, {Lindegren}, {van
  Leeuwen}, {Casta{\~n}eda}, {Fabricius}, {Bastian}, {Hambly}, {Hern{\'a}ndez},
  {Bombrun}, {Evans}, {De Angeli}, {Riello}, {Busonero}, {Crowley}, {Mora},
  {Lammers}, {Gracia}, {Portell}, {Biermann}, \& {Brown}}]{rowell2021gaia}
{Rowell}, N., {Davidson}, M., {Lindegren}, L., {et~al.} 2021, \aap, 649, A11,
  \dodoi{10.1051/0004-6361/202039448}

\bibitem[{{Spergel} {et~al.}(2015){Spergel}, {Gehrels}, {Baltay}, {Bennett},
  {Breckinridge}, {Donahue}, {Dressler}, {Gaudi}, {Greene}, {Guyon}, {Hirata},
  {Kalirai}, {Kasdin}, {Macintosh}, {Moos}, {Perlmutter}, {Postman},
  {Rauscher}, {Rhodes}, {Wang}, {Weinberg}, {Benford}, {Hudson}, {Jeong},
  {Mellier}, {Traub}, {Yamada}, {Capak}, {Colbert}, {Masters}, {Penny},
  {Savransky}, {Stern}, {Zimmerman}, {Barry}, {Bartusek}, {Carpenter}, {Cheng},
  {Content}, {Dekens}, {Demers}, {Grady}, {Jackson}, {Kuan}, {Kruk}, {Melton},
  {Nemati}, {Parvin}, {Poberezhskiy}, {Peddie}, {Ruffa}, {Wallace}, {Whipple},
  {Wollack}, \& {Zhao}}]{Spergel2015}
{Spergel}, D., {Gehrels}, N., {Baltay}, C., {et~al.} 2015, arXiv e-prints,
  arXiv:1503.03757, \dodoi{10.48550/arXiv.1503.03757}

\bibitem[{Turon {et~al.}(2005)Turon, O'Flaherty, \& Perryman}]{turon2005three}
Turon, C., O'Flaherty, K., \& Perryman, M. 2005, The Three-Dimensional Universe
  with Gaia, 576

\bibitem[{{Voggel} {et~al.}(2020){Voggel}, {Seth}, {Sand}, {Hughes}, {Strader},
  {Crnojevic}, \& {Caldwell}}]{voggel2020gaia}
{Voggel}, K.~T., {Seth}, A.~C., {Sand}, D.~J., {et~al.} 2020, \apj, 899, 140,
  \dodoi{10.3847/1538-4357/ab6f69}

\bibitem[{{Wang} {et~al.}(2014){Wang}, {Ma}, {Wu}, \& {Zhou}}]{wang2014new}
{Wang}, S., {Ma}, J., {Wu}, Z., \& {Zhou}, X. 2014, \aj, 148, 4,
  \dodoi{10.1088/0004-6256/148/1/4}

\bibitem[{{Wang} {et~al.}(2023){Wang}, {Yuan}, {Chen}, {Chen}, {Wu}, {Niu},
  {Huang}, \& {Liu}}]{Wang2023}
{Wang}, Y., {Yuan}, H., {Chen}, B., {et~al.} 2023, \apj, 954, 206,
  \dodoi{10.3847/1538-4357/ace963}

\bibitem[{{Yan} {et~al.}(2022){Yan}, {Li}, {Wang}, {Zong}, {Yuan}, {Xiang},
  {Huang}, {Xie}, {Dong}, {Yuan}, {Bi}, {Chu}, {Cui}, {Deng}, {Fu}, {Han},
  {Hou}, {Li}, {Liu}, {Liu}, {Liu}, {Luo}, {Shi}, {Wu}, {Zhang}, {Zhao}, \&
  {Zhao}}]{yan2022overview}
{Yan}, H., {Li}, H., {Wang}, S., {et~al.} 2022, The Innovation, 3, 100224,
  \dodoi{10.1016/j.xinn.2022.100224}

\bibitem[{{Zhan}(2011)}]{Zhan2011}
{Zhan}, H. 2011, Scientia Sinica Physica, Mechanica \& Astronomica, 41, 1441,
  \dodoi{10.1360/132011-961}

\end{thebibliography}
\bibliographystyle{aasjournal}

\end{document}